\documentclass[graybox, nosecnum]{svmult}

\usepackage{mathptmx}       
\usepackage{helvet}         
\usepackage{courier}        
\usepackage{type1cm}        
\usepackage{makeidx}         
\usepackage{graphicx}        
\usepackage{longtable}
\usepackage{longtable,ltcaption}                           
\usepackage{multicol}        
\usepackage{multirow}        
\usepackage[bottom]{footmisc}
\usepackage{hyperref}        
\usepackage{soul}            
\hypersetup{colorlinks=true,urlcolor=blue}
\usepackage[square,numbers]{natbib}
\makeindex             

\begin{document}
\title*{A Chronological History of X-Ray Astronomy Missions}
\author{Andrea Santangelo , Rosalia Madonia \thanks{corresponding author}, Santina Piraino}
\institute{Andrea Santangelo \at Institute of Astronomy and Astrophysics, University of T\"ubingen, Sand 1, 72076, T\"ubingen, Germany, \email{andrea.santangelo@uni-tuebingen.de}
\and Rosalia Madonia \at Institute of Astronomy and Astrophysics, University of T\"ubingen, Sand 1, 72076, T\"ubingen, Germnany, \email{rosalia.madonia@astro.uni-tuebingen.de}
\and Santina Piraino \at Institute of Astronomy and Astrophysics, University of T\"ubingen, Sand 1, 72076, T\"ubingen, Germnany, \email{santina.piraino@uni-tuebingen.de}}
%
%
\maketitle
\abstract{In this chapter we briefly review the history of X-ray Astronomy through its missions. We follow a temporal development, from the first instruments onboard rockets and balloons to the most recent and complex space missions. We intend to provide the reader with detailed information and references on the many missions and instruments that have contributed to the success of the exploration of the X-ray Universe. We have not included missions that are still operating, providing the world-wide community with high quality observations. Specific chapters for these missions are included in a dedicated section of the handbook.}
\section{Keywords} 
X-rays astronomy; X-rays Balloons; X-rays Rockets; X-ray space missions; History of X-ray astronomy. 
\section{Introduction}
\begin{figure}[t]
\begin{center}
\includegraphics[height=6.0cm,]{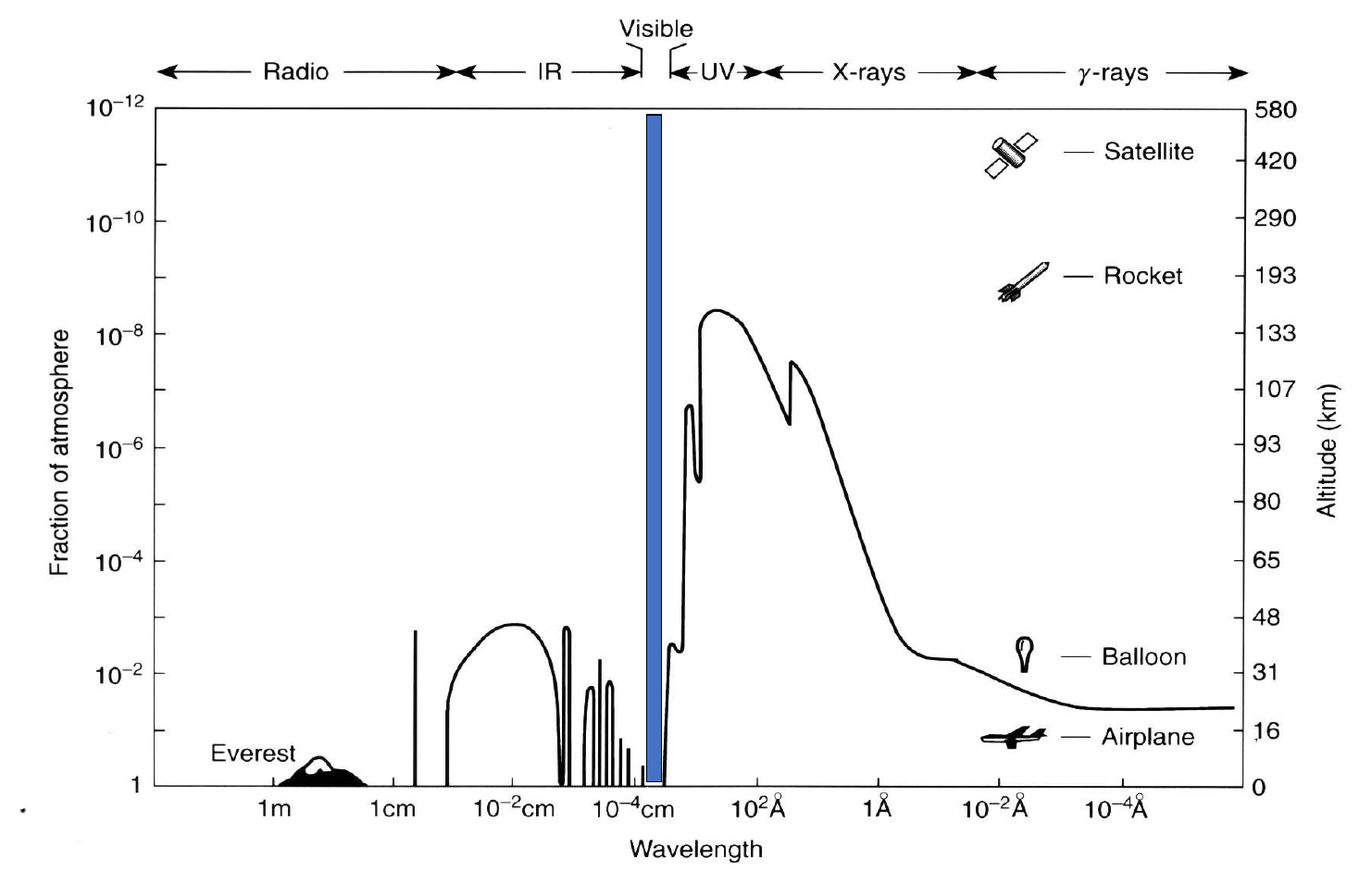}
\caption{Atmospheric absorption as a function of the wavelength (bottom axis). The solid lines indicate the fraction of the atmosphere, expressed in unit of 1 atmosphere pressure (right vertical axis) or in terms of altitude (left vertical axis), at which half of the incoming celestial radiation is absorbed by the atmosphere. Whereas radio and visible wavelength (blue rectangle) can reach without being absorbed the Earth's surface, Infrared, Ultraviolet and X-rays are strongly absorbed.
(Credit High Energy Astrophysics Group, University of T{\"u}bingen).}
\label{fig:Absorption}
\end{center}
\end{figure}

Earth's atmosphere is not (fortunately!) transparent to X-rays. In the figure published in 1968 by Riccardo Giacconi and colleagues \cite{Giacconi_et_al._1968}, a figure in many ways now historical, the attenuation of the electromagnetic radiation penetrating the atmosphere due to atmospheric absorption is presented as a function of the wavelength (see Figure \ref{fig:Absorption}). 

To explore the Universe in X-rays or in the soft gammas, it is therefore necessary to fly instrumentation onboard rockets, balloons, or satellites, and this presented new technological challenges at the end of the 1950s. The development of X-ray Astronomy therefore had to wait for the development of rockets capable of carrying instrumentation into the upper layers of the atmosphere. Its history thus coincides with the 'space race', which began after the end of World War II and experienced a decisive acceleration  with the launch of Sputnik in 1957, and Yuri Gagarin's first human space flight ever on April 12, 1961 \cite{Santangelo_Madonia_2014}. 

Following the best-known narrative, one usually traces the birth of X-ray Astronomy to the program of the AS\&E-MIT and especially to the flight of the rocket launched by Giacconi, Paolini, Rossi, and Gursky on June 18 1962 from White Sand (New Mexico), which led to the discovery of the first celestial source of X-rays, Sco X-1 \cite{Giacconi_et_al._1962}. However, the story, as we will see later, is more complex.  This paper, although detailed, is not exhaustive, further read could be found in Giacconi's book "Secrets of the hoary deep" \cite{Giacconi_2008}; in Hirsh's "Glimpsing an Invisible Universe" \cite{Hirsh_1983} and in the review chapter of Pounds "Forty years on from Aerobee 150: a personal prospective" \cite{Pounds_2002}. 

\section{The early years of X-ray astronomy}
At the beginning of the 20th century, there was great interest among scientific communities in the study of the Earth's atmosphere. The emanation power of newly discovered radioactive elements, with new types of radiation, and the discovery of cosmic rays (then called 'penetrating radiation') are probably the main reason for this interest.
Whether or not some layer of the upper atmosphere could be ionized was of particular interest for investigation\footnote{How suggested by Swann \cite{Swann_1916_a}: 
\begin{quote}
    The subject of the ionization of the upper atmosphere is one of extreme importance to students (SIC). 
    From various points of view there are indications that the upper atmosphere is to be treated as a region of high electrical conductivity.
\end{quote} He further wrote \cite{Swann_1916_b}:
\begin{quote} both of Terrestrial Magnetism and of Atmospheric Electricity, and from various points of view there are indications that this region of the atmosphere is to be treated as one of relatively high electrical conductivity.
\end{quote}}.  

Working on radio waves, Merle Tuve and Gregory Breit \cite{Tuve_and_Breit_1925} noted an interference phenomenon hypothetically due to the existence of an ionized reflecting layer in the upper atmosphere (the Kennelly-Heaviside layer or E layer) \cite{Tuve_1967}\footnote{The work of Tuve and Breit started a new research's branch that in the end brought to the invention of the Radar}.
 Between 1925 and 1930, Edward O. Hulburt published different papers on the reflecting properties of the Kennelly-Heaviside layer of the atmosphere \cite{Taylor_and_Hulburt_1926, Hulburt_1928}. He suggested that this should have been related to some Sun activity, because the ionization of the atmosphere could only be due to absorption of the Sun's ultraviolet light or, more likely, X-rays. When, immediately after World War II, the U.S. military offered research organizations and scientific institutions the opportunity to fly scientific instruments aboard V-2 rockets, developed during the war by Wernher von Braun, Edward Hulburt, head of the Naval Research Laboratory (NRL), enthusiastically accepted the offer to further investigate the reflective power of the atmosphere. Herbert Friedman was working in Hulburt's department. He was interested in studying the Sun's UV and X-rays to understand their role in the formation of the ionosphere. 
 
\begin{figure}
    \centering  
    \includegraphics[width=10cm]{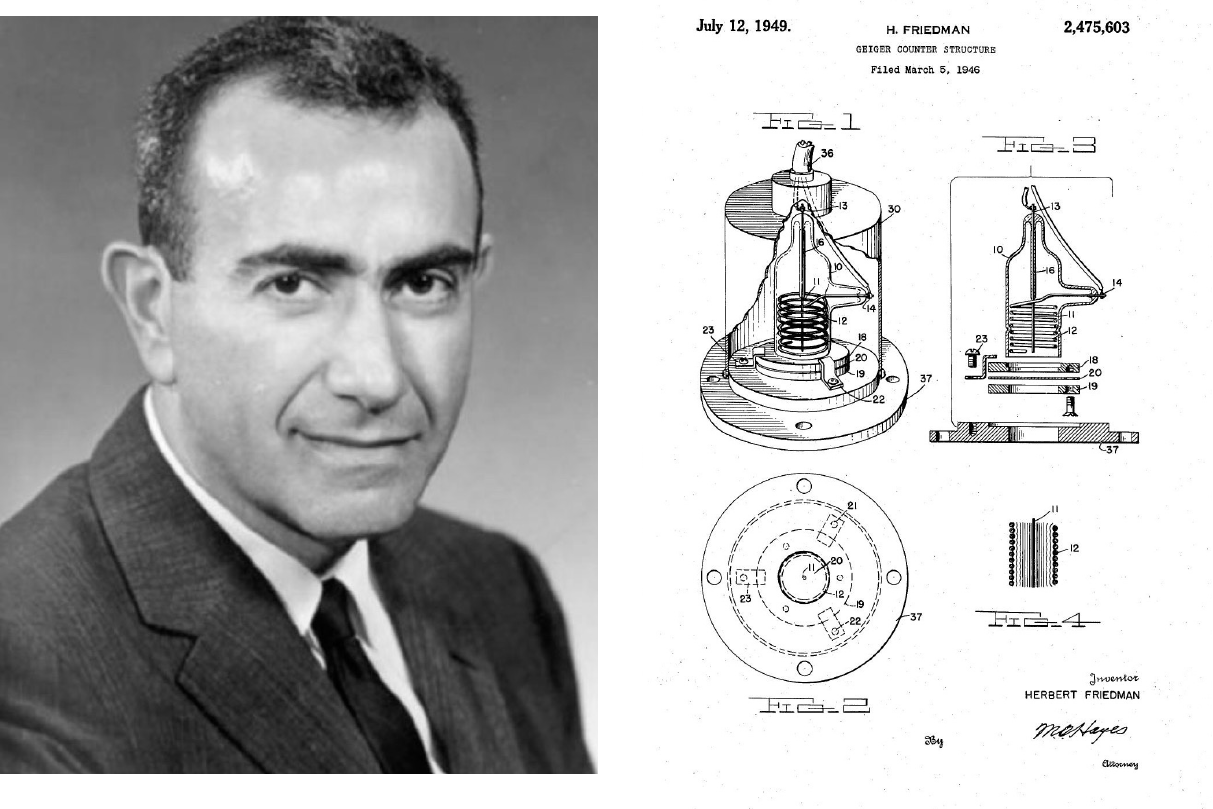}
    \caption{Left: Herbert Friedman (1916-2000) was certainly a pioneer in X-ray research of the celestial sources. Right: Friedman's U.S. Patent No. 2,475,603, for an adaptation of the tube used in a Geiger-Mueller counter. Thanks to a reduced background, Friedman's tube design increased the counter's sensitivity to weak sources. The details of the figure can be found at \cite{Friedman_Chicago_University_Library} Credit: Public Domain}
    \label{fig:Friedmann_Geiger}
\end{figure}

 Using a combination of filters and gas mixtures, Friedman built several photomultiplier tubes, each sensitive in a narrow frequency range. With the V-2 number 49 flight, launched in September 1949 from White Sands \cite{Hirsh_1980}, Friedman and colleagues confirmed the hypothesis that the ionization of the atmosphere above 87 km was due to solar X-rays emitted by the sun corona \cite{Friedman_et_al.1951}. For the first time X-ray instrumentation had been launched above the Earth's atmosphere.
 Further developments in the field were obtained thanks to the construction of a new type of rocket, the Aerobee serie, by James van Allen. Using Aerobee rockets, Friedman and colleagues conducted a series of night flights to search for stellar sources that, as the sun, could emit UV and X-ray radiation \cite{Friedman_et_al.1951,Kupperian_et_al._1959}. Only an upper limit of $10^{-8}$~ergs~cm$^{-2}$~s$^{-1}$~\AA$^{-1}$ was obtained. Herbert Friedman (see Figure \ref{fig:Friedmann_Geiger})  was a pioneer of X-ray astronomy: he obtained the first X-ray image of the sun with a pinhole camera, and flew the first Bragg spectrometer for measuring hard X rays.  The first satellite, SOLRAD, for long-term monitoring of the sun was also conceived and developed by Friedman.
 
The prewar interest in the physics of the upper atmosphere and its interaction with the solar radiation, was also strong in the UK, at the Imperial College and at the University College London (UCL)\footnote{Pioneers of ionospheric physics and geomagnetism were Sir Edward Appleton, Sydney Chapman, John Ashworth Ratcliffe,  Harrie Massey and his student David Bates, and James Sayers.}. 
In 1942, the Gassiot Committee, a committee concerned with upper atmospheric research, of the Royal Society began to cooperate with the Meteorological Research Committee in order to consider the use of rockets for a program for atmospheric physics \cite{Pounds_2010}. The war and the postwar economic difficulty of the nation slowed down the program, but eventually the common interest of military institutions and scientific groups gave the scientific community the possibility to fly instrumentation on board of rockets and later satellites. Most remarkable was the agreement between the head of the Physics department at UCL, Harrie Massey\footnote{According to Pounds "{...} Sir Harrie Massey {...} was the key player in establishing the UK as the clear leader --after the USA and the Soviet Union-- in the early years of space research \cite{Pounds_2010}.}, and Sir Arnold Hall, director of the Royal Aircraft Establishment (RAE) Farborough\footnote{We cite Ken Pounds \cite{Pounds_2010}: "[...] on 13 May 1953, when the chairman of the Gassiot Committee was about to leave for Shenley to play in the annual UCL staff-students cricket match. Massey's response to the question, 'would there be interest in using rockets available from the Ministry for scientific research?' was an immediate 'yes' [...]"}. The result was the funding of the Skylark Program and the formation of space research group at UCL (Robert Boyd), Imperial College (Percival Albert Sheppard), Birmingham University (James Sayers), Queen's University of Belfast (David Bates and Karl George Emeleus), and University College Wales, Aberystwyth (William Granville Beynon). The early Skylark missions were dedicated to the study of the sun, the Lyman-$\alpha$ and X-ray emission, as well as the study of upper atmosphere, as already said. The International Geophysical Year (1957/58) was a motivating occasion for the development of projects and studies. Indeed, the first successful test launch of a Skylark rocket occurred on the 13th of February 1957 from Woomera Range, Adelaide Australia. The choice of Woomera was due to the personal Australian relationships of H. Massey\footnote{Massey was born on May 16, 1908 in Invermay, Victoria, Australia.}. Nine months later on November 13th the first Skylark rocket with a scientific payload was launched from Woomera. The program was very successful, new instruments for X-ray were developed by a group of young scientist of the UCL, among them Ken Pounds who in 1960 received an assistant lectureship at Leicester with a three years grant of £13006 from the Department of Scientific and Industrial Research. This generous fund was indubitably a consequence of the launch of the Sputnik I.
The two new instrument developed by UCL and then the Leicester group were: a photographic emulsion protected in an armoured steel cassette with the filter mounted behind aluminum and beryllium foils\footnote{"This device was flown successfully in over 20 Skylark flights during the 1960s, providing the first direct broad band X-ray spectra over a wide range of solar activity \cite{Pounds_1986}"}; and a Proportional Counter Spectrometer (PCS) that according to Ken Pounds was to became the workhorse detector in X-ray astronomy \cite{Pounds_2010}".

The study of solar X-rays were pioneered in the Netherlands by Kees de Jager, who started a laboratory for space science at the University of Utrecht. He was supported by the atomic physicist Rolf Mewe, who developed theoretical models for data interpretation. Also the Cosmic-Ray Working Group at the University of Leiden, in the sixties, worked on X-ray with rockets and balloons in collaboration with the Nagoya University and ISAS Institute in Japan. Eventually in 1983 these groups, together with the Groningen University joined in the SRON (Stichting Ruimte Onderzoek Nederland) with the aim to develop instruments for space science missions. 

As already mentioned, the turning point for space activities in general, and therefore for X-ray astronomy, was the so called 'Sputnik shock' of 1957. New opportunities appeared and space research was welcomed and financed. In the US, in September 1959, Bruno Rossi, chairman of the board of the American Science \& Engineering (AS\&E), a startup high-tech company formed in Cambridge a year earlier by Martin Annis, suggested to Riccardo Giacconi, who was called from Princeton to become head of the Space Science Division of the AS\&E, to develop some research program on X-ray astronomy. In the next few months, on February 1960, two proposals were submitted to the newly formed NASA: one, rather visionary, to develop a X-Ray telescope (Wolter type), and another for a rocket mission to investigate the emission (or scattering) of X-Rays from the moon and from the Crab Nebula. NASA accepted the first proposal and refused the second one. In an oral interview Nancy Roman, of NASA, said that this proposal was not funded because, in her mind, it was impossible to detect such emission \cite{Roman_1980}\footnote{According to Nancy Roman: "Yes. The first X-ray work was '62, if I remember right, and that was funded by the Air Force. I didn't fund it. I guess you can blame me for being too good a scientist or you can blame me for not having foresight. Giaconni came to me with a proposal to fly an experiment to measure solar X-rays scattered off the moon, and it was, to me, absolutely clear that that was impossible. Still is.[...] If they had come to me to say they wanted to do a sky survey in X-ray, I think, admittedly in hindsight, that I would have supported them, because I was very much aware of the desirability of finding out something about new wave length regions. But I could not see supporting an experimental rocket to measure reflected solar X-rays from the moon." Note that the misspelling of the name of Giacconi is already in the original transcript.}. Nevertheless, certain of the importance of rocket mission and waiting for the realisation of the telescope, Giacconi sent the proposal to the Air Force Cambridge Research Laboratory that, on the contrary to NASA, funded a series of rocket launches. The first and second Aerobee rocket launch failed but the third one was very successful: it changed the history of astronomy and our perception of the Universe.
It is fair to mention that in December 1960, a year before the discovery of Sco X-1, Philip Fisher of the Lookheed Company, had submitted a proposal to NASA to search for cosmic X-ray sources \cite{Fisher_1960}. Nevertheless the launches of the Loockheed rockets (Aerobee 4.69 and 4.70) occurred on September 30 1962, and March 15 1963, after the discovery of Sco X-1. However, according to Fisher \cite{Fisher_2009}, his results were not properly taken into account and cited in the subsequent scientific meetings focused on X-Ray Astronomy. 

\section{Rockets and balloons in the 60s and 70s}

\subsection{Rockets}
 
\begin{figure}[t]
\begin{center}
\includegraphics[height=6.0cm]{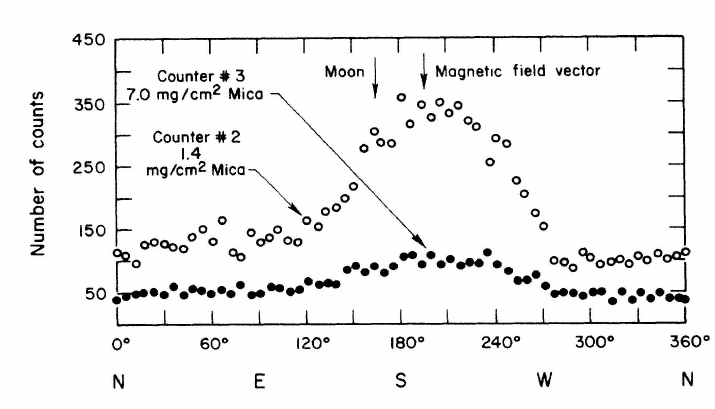}
\caption{The discovery plot which marked the beginning of X-ray astronomy. The azimuthal distribution of the count rates of the Geiger-M\"uller detectors flown in the June 1962 flight is shown. (Credit \cite{Giacconi_et_al._1962})  \label{fig_5}}
\label{fig:DiscoveryScoX1}
\end{center}
\end{figure}

On June 18, 1962 an Airforce Aerobee rocket was launched from White Sand missile range in New Mexico with the aim and the appropriate instrumentation to search for X-ray emission from celestial objects. The pioneer AS\&E-MIT experiment discovered the first extra solar source of X-rays (See Figure \ref{fig:DiscoveryScoX1}), a diffuse X-Ray background component, and the probable existence of a second source in the proximity of the Cygnus constellation \cite{Giacconi_et_al._1962}. The payload consisted of three Geiger counters, each composed of seven mica windows of 20 cm$^2$ comprising area, the detectors had a sensitivity between 2 and 8 {\AA}. 

\par

As predicted by Nancy Roman, no X-rays from the moon were observed. That was just the beginning: an intense program based on rocket launches was started. Rockets observed for only a few minutes, from a maximum altitude of $\sim 200$\,km. 
The list of the rocket experiments launched until 1970 is included in Appendix 1.

The large majority of rocket launches was performed by US scientists. However, the UK participated to the early race of X-ray astronomy, with the Skylark launches. In particular, Skylark launches SL118 and SL119, from Woomera, Australia, provided for the first time a survey of X-ray sources in the Southern Hemisphere.

\subsection{Balloons} 
The short fly-time of the sound rockets was a clear limit for the study of the X-ray sources, especially once their variability was revealed. With the use of aerostatic balloons, long-duration observations on the order of hours became possible, even if from a lower altitude. Different research groups undertook intense and fruitful balloon campaigns. In particular, the group at MIT of George W. Clark, Gordon P. Garmire and Minoru Oda, on leave from Tokyo University, was involved in a robust and successful program for X-ray sky observations with balloons. A detailed list of the MIT Balloon flights is reported in Appendix 2. The efforts to fly balloon mission for X-ray astronomy were undertaken by other institutions world-wide. In particular, the following groups were rather active: Leiden University; the GIFCO group at Bologna University and the TESRE institute of the Consiglio Nazionale delle Ricerche (CNR); the Centre d'Etudes Nucléaires (CEN) Saclay (France); the TATA institute in Mumbai (India); the Rice university Houston (Texas); the Adelaide University (Australia) and Nagoya University (Japan). A detailed list of the Balloon flights launched by these institutions is reported in Appendix 3. 

We wish to mention in particular that a balloon program in the hard X-rays (20-200 keV) was pursued by the Institut f\"ur Astronomie und Astrophysik der Universit\"at T\"ubingen (AIT) and the Max Planck-Institut f\"ur Extraterrestrische Physics in Garching (MPE) from 1973 to 1980 with nine successful balloon flights from Texas and Australia \cite{Staubert_et_al._1981}. The program was led by Joachim Tr\"umper who started German X-ray astronomy in T\"ubingen before moving to the directorship at MPE. A detailed list of the Balloon flights launched by these institutions is reported in Appendix 4. The instruments were built and operated by MPE and AIT and consisted of scintillation counters with NaI(Tl)-crystals \cite{Kendziorra_et_al._1974, Reppin_et_al._1978}. 
The close cooperation between AIT and MPI continued in the 80s with the construction and operation of the High Energy X-Ray Experiment (HEXE) used during three successful balloon campaigns (May 1980 and September 1981 launched from Palestine/Texas, as well as November 1982 launched from Uberaba/Brazil). 

At the beginning of the 1970s, the main world wide available balloon launch site was the NSBF facility in Palestine, Texas, USA. In the period between 1967 to 1976, the average flight duration was about 10-15 hours, with a few exceptions (4 flights lasted 40-60 hours and only one, in 1974, up to 120 hours). 
The opening in 1975 of the Stratospheric Balloon Launch Base of Trapani-Milo in Sicily provided the opportunity to use transatlantic flights whith long and stable durations, and the capability to carry payloads with mass up to 2--3 tons at altitudes above 38--42~km, perfect to realize X-Ray investigations of cosmic sources \cite{Ubertini_2008}. The flight campaign started with a precursor flight operated by the Italian Trapani-Milo ground operation crew and a launch team from NSBF-NASA. The payload had a total weight of 1500 kg out of which 500~kg of scientific experiments and flight services. The flight started on August 5, 1975, and safely landed on the US east coast after a flight of 81 hours. Several successful balloon experiment were performed  by the Istituto di Astrofisica Spaziale of Rome (IAS), in collaboration with others Italian and international institutes\footnote{Among them the Istituto di Fisica Cosmica of Milan (IFC), the Laboratorio di Tecnologie e Studio delle Radiazioni Extraterrestri of Bologna (TESRE), the Istituto di Fisica Cosmica con Applicazione all'Informatica  of Palermo (IFCAI), and international Institutions such as University of Southampton and RAL (UK), TATA Institute (Mumbai, India), T{\"u}bingen University (Germany), ADFA (Australia), CNES (France), INTA (Spain), etc \cite{Ubertini_2008}}. 
The list of the major transatlantic Balloon missions is reported in Appendix 4.





\section{Uhuru and the others, opening the age of the satellites in the early 70s}

The first satellite designed for cosmic X-ray observation was the US Vanguard 3 satellite, launched on September 18, 1959. It operated until December 11, 1959. The payload consisted of ion chambers provided by NRL that were intended to detect (solar) X-rays (and Lyman-alpha). Unfortunately as noted in \cite{Friedman_1960} "the Van Allen Belt radiation swamped the detectors most of the time and no useful X-ray data were obtained". On October 13, 1959 the US Explorer 7 satellite was launched from Cape Canaveral. It operated until August 24, 1961, and, like Vanguard 3, carried, among other experiments, ion chambers provided by NRL. The goal was to detect (solar) X-rays (and Lyman-alpha). Unfortunately, no useful X-ray data were obtained similar to the cas of Vanguard 3 \cite{Friedman_1960}. 

 \subsection{\textsc{Uhuru}}

\begin{figure}[h]
\begin{center}
\includegraphics[height=5.0cm]{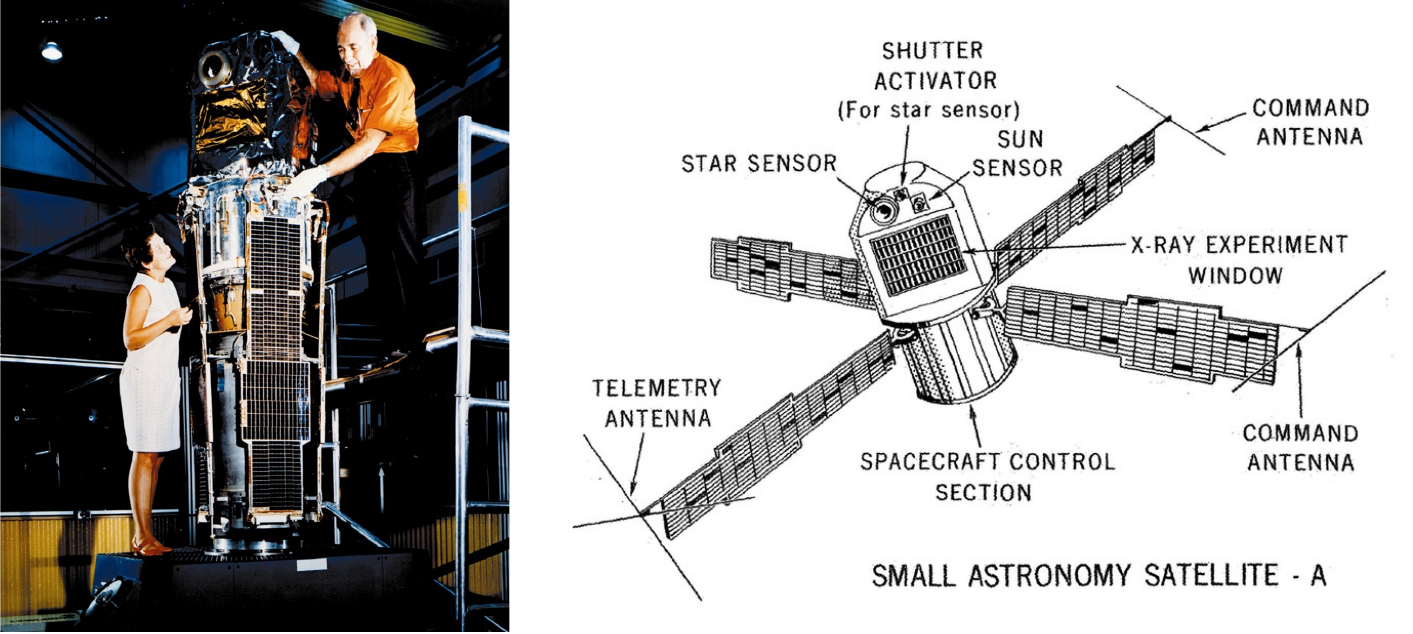}
\caption{Left: Marjorie Townsend discusses the SAS-1 X-ray Explorer Satellite's performance with Bruno Rossi during preflight tests at NASA's Goddard Space Flight Center. Marjorie Townsend was the first woman to become a satellite project manager at NASA. Right: a schematic of the satellite. All major basic elements of a X-ray satellite are shown. (Credit NASA)} 
\label{fig:Uhuru}
\end{center}
\end{figure}

The Small Astronomical Satellite 1 (SAS-1) was the first of small astronomy satellites developed by NASA, and was entirely devoted to the observations of cosmic X-ray sources (see Figure \ref{fig:Uhuru}). It was launched on December 12, 1970 from the Italian San Marco launch platform off the coast of Kenya, operated by the Italian Centro Ricerche Aerospaziali. December 12 was the seventh anniversary of the independence of Kenya, and in recognition of the kind hospitality of the Kenyan people, Marjorie Townsend, the NASA mission project manager, named the successfully launched mission 'Uhuru', Swahili for 'freedom'. 
Uhuru was launched into a nearly equatorial circular orbit of about 560 km apogee and 520 km perigee, with an inclination of 3° and an orbital period of 96 minutes. The mission ended in March 1973. The X-ray detectors consisted of two sets of large-area proportional counters sensitive (with more than 10 percent efficiency) to X-ray photons in the 1.7-18\,keV range. The lower limit was determined by the attenuation of the beryllium windows of the counter plus a thin thermal shroud, needed to maintain the temperature stability of the spacecraft. The upper energy limit was determined by the transmission properties of the filling gas. Pulse-shape discrimination and anticoincidence techniques were used to reduce the particle background and high-energy photons \cite{Giacconi_et_al._1971}. The main features of the mission are reported in Table \ref{tab:Uhuru}.

\begin{table}[h]%
\caption{Uhuru\label{tab:Uhuru}}
\centering
\begin{tabular}{ p{4.5cm} p{3.5cm} p{3.5cm}} 
\hline
      
\hline
Instrument & Set 1  & Set 2 \\
\hline
Bandpass~(keV) &1.7-18 &1.7-18 \\
Eff Area~(cm$^2$) &840&840 \\
Field of view (FWHM)& 0.52$^{\circ}$$\times$ 5.2$^{\circ}$&5.2$^{\circ}$ $\times$ 5.2$^{\circ}$\\
Timing resolution~(s) &0.192 &0.384\\
\hline
Sensitivity (ergs cm$^{-2}$ sec$^{-1}$)& \multicolumn{2}{l}{\hspace{1.5cm}1.5$\times$10$^{-11}$}\\
\hline
\end{tabular}
\end{table}

\begin{figure}[h]
\begin{center}
\includegraphics[height=6.0cm]{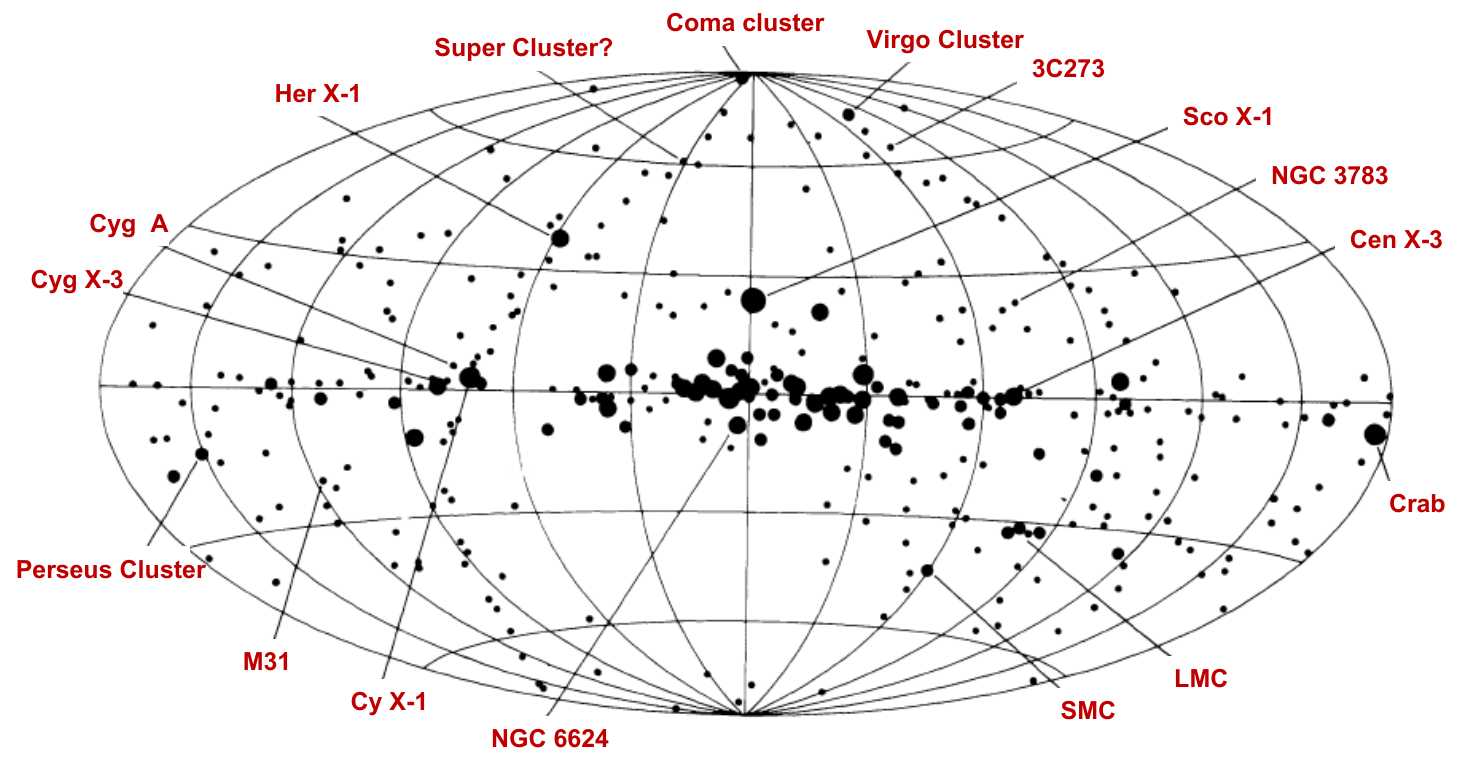}
\caption{The Map of the X-ray sky after Uhuru, according to the fourth Uhuru catalogue. Figure adapted from \cite{Forman_et_al_1978}.} 
\label{fig:Uhuru_survey}
\end{center}
\end{figure}

The main science achievement of Uhuru was, with no doubt, the completion of the first X-ray all sky survey up to a sensitivity of 0.5 mCrab (between 10 and 100 times better  than what achievable with rockets). Uhuru detected 339 X-ray sources of different classes: X-ray binaries, supernova remnants, Seyfert galaxies, and clusters of galaxies, for which diffuse X-ray emission was discovered \cite{Forman_et_al_1978} (Figure \ref{fig:Uhuru_survey}). 

\subsection{\textsc{Apollo 15 and Apollo 16}}
On July 26, 1971, the Apollo 15 lunar mission carried,  inside  the Scientific Instrument Module (SIM) of the  Service Module, an X-ray fluorescence spectrometer (XRFS \cite{Jagoda_et_al._1974, Adler_et_al._1975}) and a gamma-ray spectrometer (GRS), with the aim to study the composition of the lunar surface. Similarly, on April 16, 1972 the same suite of instruments was flown on Apollo 16. The XRFS was manufactured by the AS$\&$E. The main objective was indeed to study the Moon's surface from lunar orbit, in order to better understand the Moon's overall chemical composition (see \cite{Gloudemans_et_al._2021}). On the way back from the Moon to the Earth (i.e., during the ‘trans-Earth coast’) the XRFS observed parts of the X-ray sky. The prime objective of the Apollo observations during the trans-Earth coast was to understand the nature of the X-ray sources discovered earlier (e.g., Cyg X-1, Sco X-1) by observing them continuously for approximately half an hour to an hour, which was unique at that time. UHURU could only observe for approximately 1 or 2 min per sighting. Preliminary results were reported in the Apollo 15 and  16 Preliminary Science Reports \cite{Adler_et_al._1972a, Adler_et_al._1972b}. Further results from the trans-Earth coast observations include a mysterious (Type I?) burst seen by Apollo 15 (see \cite{Kuulkers_et_al._2009}) and a gamma-ray burst seen by Apollo 16 \cite{Metzger_et_al._1974, Trombka_et_al._1974}.
\begin{figure}[h]
\begin{center}
\includegraphics[height=5.0cm]{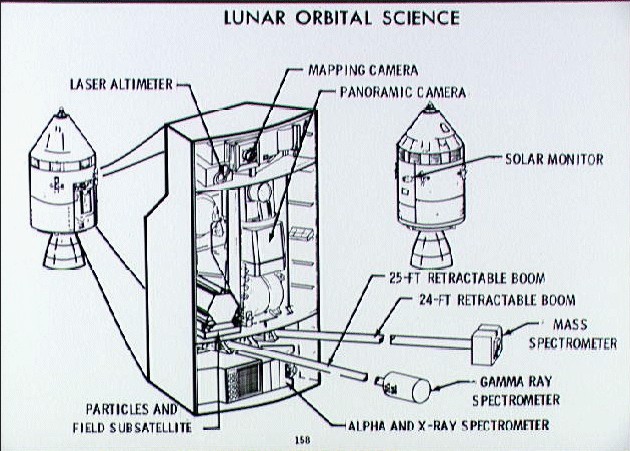}\includegraphics[height=5.0cm]{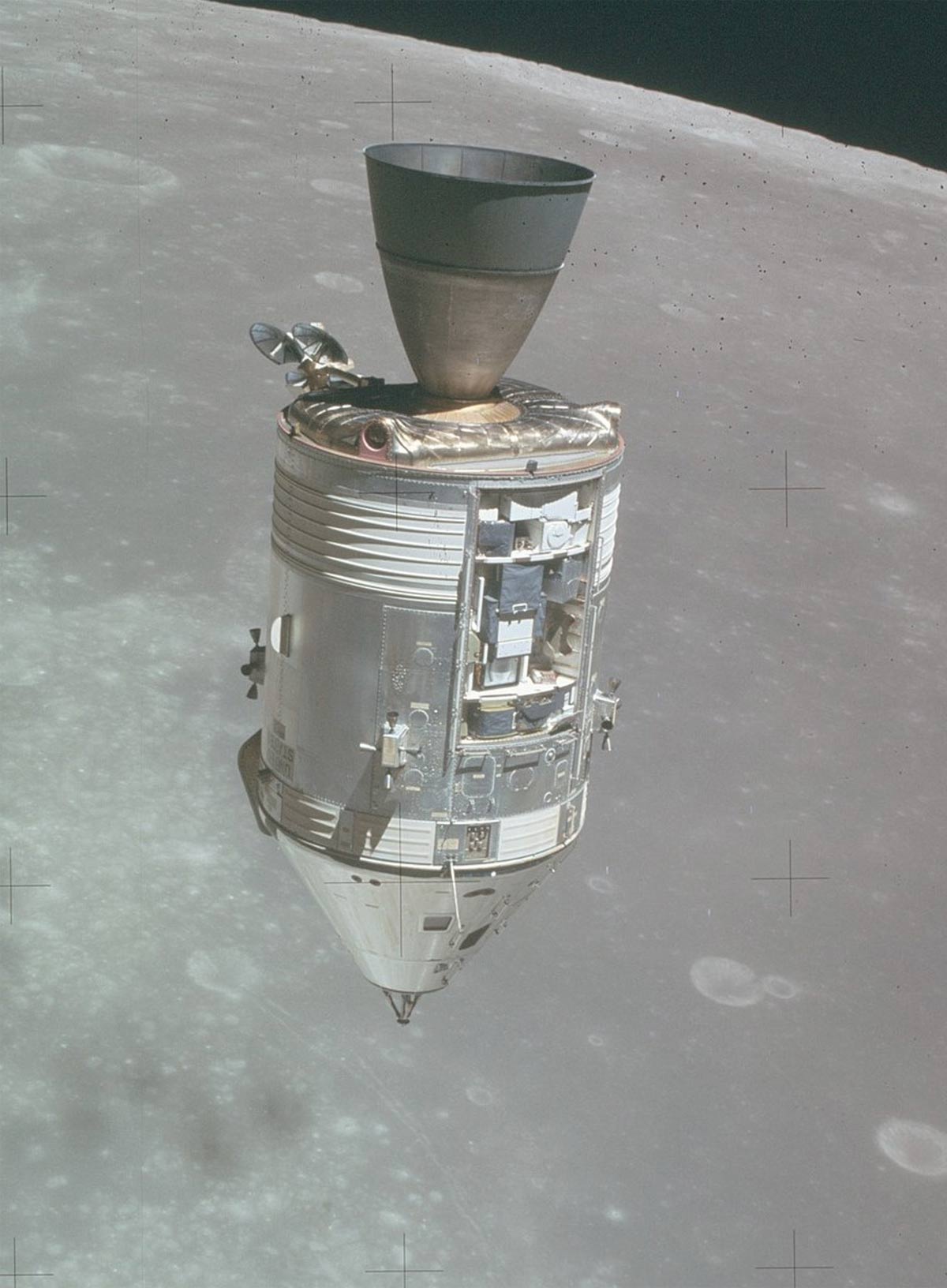}
\caption{The Scientific Instrument Module of Apollo 15 (Credit HEAG@UCSD \cite{ucsd}).} 
\label{fig:Apollo_LOS}
\end{center}
\end{figure}
 \subsection{\textsc{SAS-3}}
\begin{figure}[h]
\begin{center}
\includegraphics[height=4.5cm]{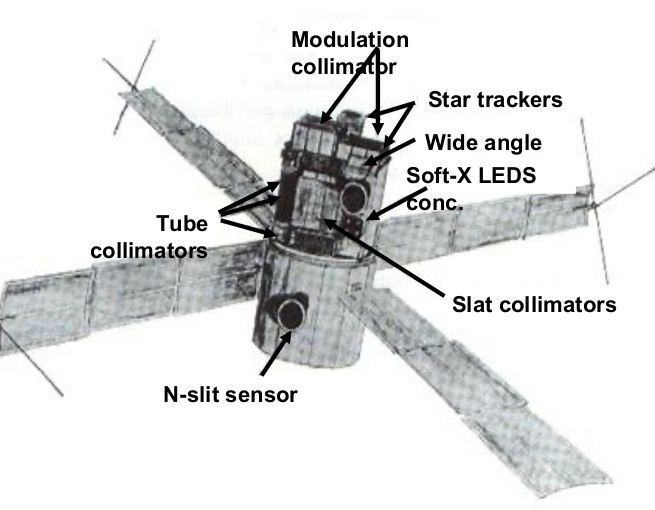}\includegraphics[height=4.5cm]{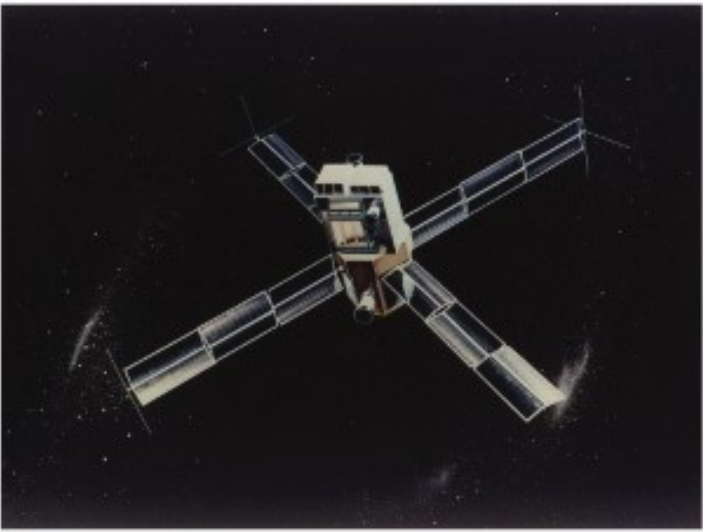}
\caption{Left: a schematic diagram of the instruments of the science payload. SAS3 was already a complex mission. Note that it also had onboard a set of 4 grazing incidence concentrators.  Right: an artistic impression of the SAS-3 satellite. (Credit NASA)} 
\label{fig:SAS3}
\end{center}
\end{figure}

The second
small satellite for X-ray astronomy SAS-3 was launched on May 7, 1975 again from the Italian San Marco launch facility. Its initial orbit was equatorial. SAS-3 was designed as a spinning satellite. The spin rate was controlled by a gyroscope that could be commanded to stop rotation. In this way, all instruments could be pointed providing about 30 minutes of continuous exposures on sources such as a pulsars, bursters, or transients. The nominal spin period was 95 minutes, which was also the orbital period having an inclination of 3$^{\circ}$ and an altitude of 513~km. The scientific payload \cite{Mayer_1975}, designed and built at MIT, consisted of four X-ray instruments (see Figure \ref{fig:SAS3}):

\begin{itemize}
     \item 2 rotating modulation collimator systems (RMCS \cite{Schnopper_et_al._1976}), each of which had an effective area of 178\,cm$^2$, and consisted of a modulation collimator and proportional counters active in the energy bands of 2-6 and 6-11\,keV. The collimator had an overall FOV of 12$^{\circ}\times$12$ ^{\circ}$, with a FWHM of 4.5 arcmin, centered on the direction parallel to the spin axis (satellite +Z-axis).
    
      \item  3 crossed slat collimators (SME \cite{Buff_et_al._1977}) each with a proportional counter. They were designed to monitor a large portion of the sky in a wide band of directions centered on the plane perpendicular to the rotation axis of the satellite. Each detector had an on-axis effective area of 75 cm$^2$. 
      The collimators defined 3 long, narrow FoVs which intersected on the +Y axis and were inclined with respect to the YZ plane of the satellite at angles of -30$^{\circ}$, 0$^{\circ}$, and +30$^{\circ}$, respectively. During the scanning mode, an X-ray source would appear in the 3 detectors. Three lines could then be obtained, and their intersection determined the source position. The central collimator had a field of view of 2$^{\circ}\times$120$^{\circ}$ with FWHM 1$^{\circ}\times$32$^{\circ}$. The left and right collimators had narrower, but similar responses, i.e., 0.5$^{\circ}\times$32$^{\circ}$ (FWHM) and 1.0$^{\circ}\times$100$^{\circ}$ (FW). The proportional counters were filled with argon and were sensitive in the range 5-15\,keV. In addition the central detector featured a xenon counter, located behind the argon detector, that extended the energy range to 60\,keV. Over the energy range 1.5-6\,keV, 1 count/s was equivalent to $1.5 \times 10^{-10}$erg cm$^{-2}$ sec$^{-1}$ for a Crab-like spectrum. In any given orbit, at the nominal 95 min spin period, 60\% of the sky was scanned by the center slat detector with an effective area from 300-1125 cm$^2$.
      \item  3 tube collimators (TC \cite{Lewin_et_al._1976a}), sensitive to X-rays in the range 0.4-55\,keV, located above and below, each of which with an effective area of 80 cm$^2$. The third was along 'the left' with an effective area of 115 cm$^2$ of the slat collimators, that defined a $1.7^{\circ}$ circular FOV. The tube collimator above the slat collimator was inclined at an angle of 5 degrees above the Y-axis, and could therefore be used as a background reference for the other tube collimators aligned along the Y-axis. 
      \item  1 low-energy detector system (LEDS \cite{Hearn_et_al._1976}) to the right of the slat collimators. It consisted of a set of 4 grazing incidence, parabolic reflection concentrators with 2 independent gas-flow counters sensitive to X-rays in the range 0.15-1.0\,keV, and with an effective area of 20 cm$^2$. 
\end{itemize}

The major scientific objectives were reaching a position accuracy of bright X-ray sources to $\sim15$ arcs; study of a selected sample of sources over the energy range 0.1-55\,keV and search the sky for X-ray novae, flares, and other transient phenomena. The science highlights of the mission included the discovery of a dozen X-ray burst sources \cite{Lewin_et_al._1976a}, among which include the Rapid Burster \cite{Marshall_et_al._1979} the first discovery of X-ray from an highly magnetic White Dwarf (WD) binary system, AM Her \cite{Hearn_and_Richardson._1977}, the discovery of X-ray from Algol and HZ 43 \cite{Schnopper_et_al._1976}, the precise location of about 60 X-ray sources and the survey of the Soft X-ray background (0.1-0.28 kev) \cite{Marshall_and_Clark_1984}.

\subsection{\textsc{Heao-1}}

In 1977, NASA started launching a series of very large scientific payloads called High Energy Astronomy Observatories (HEAO). They were launched by Atlas Centaur rockets. The payloads were about 2.5\,m$\times$5.8\,m in size and $\sim$3 000\,kg in mass \cite{Bradt_1992,Tucker_1984,Peterson_1975}. The telemetry rate was large, $\sim$6 400 bits/s compared to the 1 000 bits/s typical of earlier satellites. The first of these missions, HEAO-1 (HEAO-A before launch) surveyed the X-ray sky almost three times over the 0.2 keV-10\,MeV energy band, and provided nearly constant monitoring of X-ray sources near the ecliptic poles. More detailed studies of a number of objects were made through pointed observations lasting typically 3-6 hours.

HEAO-1, operated from August 12, 1977 to January 9, 1979 in a satellite orbit at 432 km, with $23^{\circ}$ inclination and a period of 93.5\,min. The science payload included four major instruments (for the details see Table \ref{tab:HEAO1}): 

\begin{itemize}
 
\item  A1 - a Large Area Sky Survey experiment (LASS) consisting of a proportional-counter array (seven modules), sensitive in the 0.25-25\,keV energy range, designed to survey the sky for discrete sources \cite{Friedman_1979}
\item A2 -  a smaller proportional-counter array, the Cosmic X-ray Experiment (CXE), designed principally to study the diffuse X-ray background from 0.215-60\,keV \cite{Rothschild_1979,Boldt_1987}. It consisted of six proportional counters:
\begin{itemize}
\item Low Energy Detectors (LED), 2 detectors operating in the 0.15-3.0\,keV energy range
\item Medium Energy Detector (MED) operating in the 1.5-20\,keV range
\item High Energy Detector (HED), 3 detectors in the 2.5-60\,keV energy range
\end{itemize}

\item A3 - a Modulation Collimator (MC) experiment, covering the energy range 0.9-13.3 keV, with 2 detectors (MC1and MC2). It was designed to determine accurate ($\sim$1') celestial positions \cite{Gursky_1978}, 
\item A4 - a high-energy experiment, the Hard X-Ray/Low Energy Gamma Ray Experiment \cite{Matteson_1978,Peterson_1975}, extending to $\sim$10\,MeV, consisting in seven inorganic phoswich scintillator detectors:
\begin{itemize}
\item Low Energy Detectors, 2 detectors in the 15-200\,keV range 
\item Medium Energy Detectors operating in the 80\,keV-2\,MeV range
\item High Energy Detector in the range 120 keV-10\,MeV 
\end{itemize}
\end{itemize}

\begin{table}[h]%
\caption{HEAO-1 payload.\label{tab:HEAO1}}
\centering
\begin{tabular}{ p{1.8cm} p{1.9cm} p{1.cm} p{1.cm} p{0.9cm} p{0.9cm} p{1.0cm} p{1.1cm} p{1.0cm} p{1.3cm}} 
\hline
\hline
Payload & A1(LASS) &   \multicolumn{3}{c}{A2(CXE)} & \multicolumn{2}{c}{\hspace{-0.4cm}A3(MC)} & \multicolumn{3}{c}{A4} \\

Detector&  &LED&MED&HED &MC1&MC2& LED&MED&HED \\
\hline

Energy Range~(keV) &0.25-25 & 0.15-3&1.5-20&2.5-60&\multicolumn{2}{c}{\hspace{-0.4cm}0.9-13.3}&15-200&80-2000&120-10000  \\

Eff Area~(cm$^2$) & 1350-1900&2$\times$400& 800&3$\times$800& 2$\times$400& 300&2$\times$100& 4$\times$45&100 \\

\multirow{3}{*}{FOV} &\multirow{3}{*}{1$^{\circ}$$\times$4$^{\circ}$-1$^{\circ}$$\times$0.5$^{\circ}$} & &\multicolumn{2}{c}{\hspace{-0.4cm}1.5$^{\circ}$$\times$3$^{\circ}$} & & & \multirow{3}{*}{1.7$^{\circ}$$\times$20$^{\circ}$}&  \multirow{3}{*}{~17$^{\circ}$}& \multirow{3}{*}{37$^{\circ}$}\\

&&&\multicolumn{2}{c}{\hspace{-0.4cm}3$^{\circ}$$\times$3$^{\circ}$}&\multicolumn{2}{c}{\hspace{-0.4cm}4$^{\circ}$$\times$4$^{\circ}$}& && \\
&&&\multicolumn{2}{c}{\hspace{-0.4cm}3$^{\circ}$$\times$6$^{\circ}$}&&&&& \\

\hline
\hline
\end{tabular}
\end{table}

\begin{figure}[h]
\begin{center}
\includegraphics[height=6.2cm]{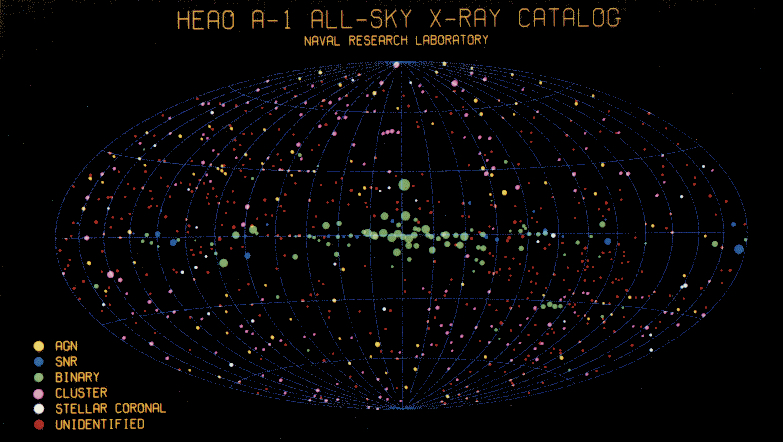}
\caption{The HEAO 1 A-1 X-Ray Source Catalog includes results from the first six months of data from HEAO-1, during which time a scan of the entire sky was completed. It contains positions and intensities for 842 sources. Half of the sources remained unidentified at the time of catalogue publication (1984). (Credit NASA).} 
\label{fig:HEAO1}
\end{center}
\end{figure}

Comprehensive catalogs of X-ray sources (one for each experiment) were obtained (see Figure \ref{fig:HEAO1}).
The large area of LASS, and the occasional pointed mode, with 1$^{\circ}\times 4^{\circ}$
FWHM collimation, enabled studies of rapid temporal variability, with e.g., the discovery of aperiodic variability in Cyg X-1 down to a time scale of 3\,ms \cite{Meekins_1984}, discovered the first eclipse in a low-mass binary system (X1658-298) \cite{Cominsky_Wood_1984,Cominsky_Wood_1989}, the 5-Hz quasi-periodic oscillation (QPO) in the 'normal-branch' mode of Cyg X-2 \cite{Norris_Wood_1987} and variability on the time scale of tens of milliseconds in an X-ray burst \cite{Hoffman_1979}.
The CXE experiment provided a complete flux-limited High Galactic Latitude
Survey (85 sources) which yielded improved X-ray luminosity functions
for active galactic nuclei and clusters of galaxies \cite{Piccinotti_1982}, a classification among AGN types \cite{Mushotzky_1984},  a  measurement of the diffuse  X-ray background from 3-50\,keV \cite{Marshall_1980,Boldt_1987}. 
The celestial positions, accurate to about 1 arcmin, obtained with the MC experiment, led to several hundred optical identifications and source classifications.
The results from the high-energy instrument included the observation of the high-energy spectra of AGN, which were key for understanding the origin of the diffuse background \cite{Rothschild_1983}, the discovery of the binary system, LMC X-4, with $\sim30$\,d periodic on-off states, and the second example (after Her X-1) of cyclotron absorption in a binary system, 4U0115+63 \cite{Wheaton_1979}.

\section{Late 1970s and the 1980s: the program in the US}

Thanks to Uhuru and HEAO-1, a new sky had been revealed and X-ray astronomy entered a new mature phase thanks to collimators and proportional counters. A key step forward was now necessary: X-ray focusing. A step that had been prepared by Riccardo Giacconi since the beginning of the 1960s, with robust R\&D plans.  

 \subsection{\textsc{Einstein}}
\begin{figure}[h]
\begin{center}
.\includegraphics[height=4.5cm]{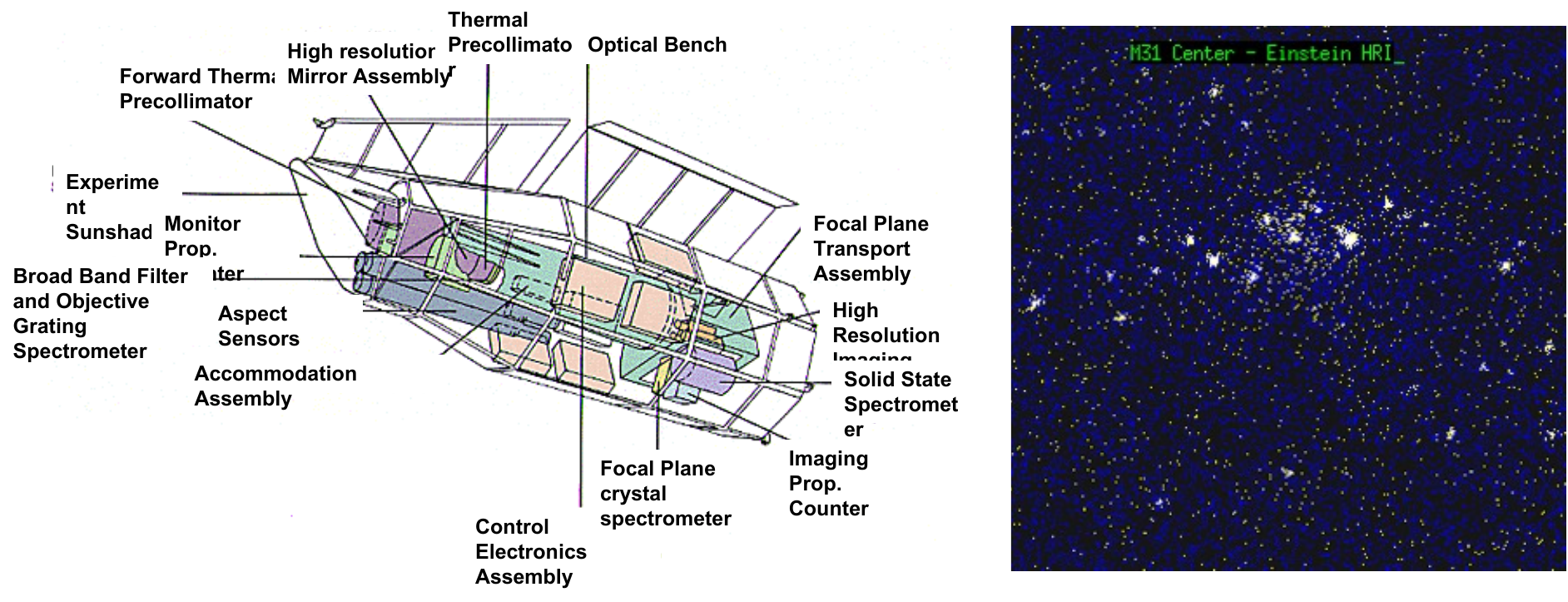}
\caption{Left: A schematic view of the Einstein satellite. Right: The Einstein view of the galactic center of the Andromeda Galaxy (M31). The power of focusing appears in the many point sources resolved (both figures credit NASA).} 
\label{fig:Einstein}
\end{center}
\end{figure}

All efforts to develop X-ray focusing telescopes resulted in a proposal to NASA for a focused Large Orbiting X-Ray Telescope (LOXT), whose team was assembled by Giacconi in 1970. Indeed, the second of NASA’s three High Energy Astrophysical Observatories, HEAO-2, renamed Einstein after launch, revolutionized X-ray astronomy thanks to its Wolter Type-I grazing-incidence X-ray focusing optics \cite{Wolter_1952} (see Figure \ref{fig:Einstein}). It was the first high-resolution imaging X-ray telescope launched into space \cite{Giacconi_1979}. Focusing enabled not only a much better position constraint, but also was key to dramatically reduce the particle background, since the volume of the detector was now significantly smaller than before. The HEAO-2 sensitivity was then several hundred times better than any previous mission. Thanks to its few arcsec angular resolution, tens of arcmin field-of-view, and greater sensitivity, it was now possible to study the diffuse emission, to image extended objects, and detect a large number of faint sources. It was a revolutionary mission in X-ray astronomy, and its scientific outcome completely changed our view of the X-ray sky. Einstein operated from November 12, 1978 to April 26, 1981, in a satellite orbit at 465-476 km, with 23.5$^\circ$ inclination and a period of 94 minutes. The scientific payload consisted of four instruments covering the energy range 0.2-20\,keV, which could be rotated, one at a time, into the focal plane of the Optics (see Table \ref{tab:einstein} for the details of the instrument parameters): 
\begin{itemize}
 
\item 
an Imaging Proportional Counter (IPC \cite{Gorestein_1981,Giacconi_1979}), operating in the 0.4-4.0\,keV with high sensitivity
\item
a High Resolution Imager  (HRI \cite{Grindlay_1980}) operating in the 0.15-3.0\,keV range 
\item a Solid State Spectrometer (SSSn \cite{Holt_1976}) in the 0.5-4.5\,keV range with moderate sensitivity
\item
a Focal Plane Crystal Spectrometer (FPCS \cite{Lum_1992}) in the 0.42-2.6\,keV range 
with very high spectral resolution $\frac{E}{\Delta E}$ of 50-100 for E$<$0.4 keV, $\frac{E}{\Delta E}$ of 100-1000 for E$>$0.4 keV).
\end{itemize}
Einstein also carried a non-focusing Monitor Proportional Counter array (MPC, \cite{Gaillardetz_1978})  
to measure the higher-energy emission (2- 15 keV) of bright sources in the view direction of the main telescopes, and an Objective Grating Spectrometer (OGS \cite{Harris_1984}), with 500~mm$^{-1}$ \& 1000~mm$^{-1}$, energy resolution $\frac{E}{\Delta E} \sim$50 was used in conjunction with HRI.

Many fundamental and far reaching results were obtained \cite{nasa_einstein}:
The high spatial resolution morphological studies of supernova remnants; the many faint sources resolved in M31 and the Magellanic Clouds; 
the first study of the X-ray emission from the hot intra-cluster medium in clusters of galaxies revealing cooling inflow and cluster evolution; the discovery of X-ray jets from Cen A and M87 aligned with radio jets;
the First medium and Deep X-ray surveys. On top of this, Einstein discovered thousands of “serendipitous” sources. 
Einstein was also the first X-ray NASA mission to have a Guest Observer program.
\begin{table}[h]%
\caption{HEAO 2 science payload \label{tab:einstein}}
\centering
\begin{tabular}{ p{2.6cm}  p{1.1cm} p{1.1cm} p{1.1cm}  p{1.8cm} p{2.cm} p{1.5cm}} 
\hline

Payload & \multicolumn{4}{c}{WolterType 1}& \hspace{0.4cm}MPC &  OGS   \\
& \textsc{IPC}& \textsc{HRI}&\textsc{SSS}& \textsc{FPCS} & & \\
\hline
Bandpass~(keV) &$0.4-4$  &$0.15-3$&$0.5-4.5$&$0.42-2.6$&\hspace{0.5cm}$1.5-20$&  \\

Eff Area~(cm$^2$) &100&5-20&200&0.1-1&\hspace{0.5cm}667& \\

\multirow{4}{*}{Field of view (FOV)}&\multirow{4}{*}{$75^\prime$}&\multirow{4}{*}{$25^\prime$}& \multirow{4}{*}{$6^\prime$}&$6^\prime$&\hspace{0.5cm}\multirow{4}{*}{$1.5^{\circ}$}&\\
& && & $1^\prime\times20^\prime$& &\\
& && & $2^\prime\times20^\prime$&&\\
& && & $3^\prime\times30^\prime$&&\\
Spatial resolution&$\sim1^\prime$&$\sim2^{\prime\prime}$&&&&\\
\multirow{2}{*}{$\frac{E}{\Delta E}$}&&&\multirow{2}{*}{$3-25$}&$50-100^{*}$  &&\\
&&&&$100-1000^{**}$&\hspace{0.4cm}$\sim20\%$ &$\sim50$ \\
\hline
\hline
\end{tabular}
\end{table}



\section{Late 1970s and the 1980s: the program in Europe}
\subsection{\textsc{Copernicus}}
Copernicus or Orbiting Astronomical Observatory 3 (OAO-3) was a collaborative effort between the USA (NASA) and the UK (SERC). The main instrument on board was the the Princeton University UV telescope (PEP) consisting of a Cassegrain telescope with an 80 cm primary mirror, a 7.5 cm secondary, and a Paschen-Runge spectrometer. In addition, the mission carried an X-ray astronomy instrument developed by the Mullard Space Science Laboratory (MSSL) of UCL. OAO-3 was launched on August 21, 1972 into a circular orbit of 7 123 km radius and an inclination of 35$^\circ$. Although some of the instruments ceased to work, it operated for almost nine years until February 1981. The X-ray Experiment (UCLXE) consisted of 4 co-aligned X-ray detectors observing in the energy-range 0.7-10\,keV: the collimated proportional counter (CPC), and 3 Wolter type-0 grazing incidence telescopes (WT-0). At the focus of the telescopes 2 proportional counters (PC1, PC2) and 1 channel photo-multiplier (CHP) were used. In Table \ref{tab:Copernicus} we report the main parameters of the instruments \cite{Bowles_et_al._1974}. Science highlights of the mission were: the discovery of several long period pulsars (e.g. X Per) \cite{White_et_al._1976}; the discovery of absorption dips in Cyg X-1 \cite{Mason_et_al._1974}; the Long-term monitoring of pulsars and other bright X-ray binaries \cite{Branduardi_Mason_Sanford_1978}; the observed rapid intensity variability from Cen A \cite{Davison_et_al.1975}. 

\begin{table}[h]%
\caption{Aside an instrument for UV astronomy, COPERNICUS carried on-board 4 X-ray detectors. \label{tab:Copernicus}}
\centering
\begin{tabular}{ l c c c c} 

\hline

\multirow{2}{*}{Instrument}& \multirow{2}{*}{CPC}&\multicolumn{3}{c}{WT-0}   \\
&  & \textsc{pc1}&\textsc{pc2} & \textsc{chp}  \\

\hline
Bandpass (nm) &$0.1-0.3$  &$0.3-0.9$&$0.6-1.8$&$2-7$  \\

Eff Area (cm$^2$) &17.8&5.5&12.5&22.9 \\
\multirow{3}{*}{FOV (FWHM)}&\multirow{3}{*}{--}&1$^{\prime}$&2$^{\prime}$&\multirow{3}{*}{10$^{\prime}$}\\
&&3$^{\prime}$&6$^{\prime}$&\\
&&10$^{\prime}$&10$^{\prime}$&\\
\hline
Energy range (keV)&\multicolumn{4}{c}{0.7-10}\\
\hline

\end{tabular}
\end{table}

\subsection{\textsc{ans}}
 ANS (Astronomische Nederlandse Satelliet) was a collaboration between the Netherlands Institute for Space Research (NIRV) and NASA. Launched on August 30, 1974, the mission re-entered the atmosphere on June 14, 1977. Its orbit should have been circular with a radius of 500 km, but due to a failure of the first-stage guidance, the final orbit was highly inclined (98$^\circ$) and elliptic (258 km perigee and 1173 km apogee) with a period $\sim$99 min. ANS took on-board three instruments: An ultraviolet telescope spectrometer (UVT) \cite{vanDuinen_et_al._1975} by the University of Groningen; a soft X-ray experiment (SXX), \cite{denBoggede_et_al._1975} developed by the University of Utrecht, that consisted of two parts known as Utrecht soft and medium X-ray detectors; and a hard X-ray experiment (HXX) \cite{Gursky_et_al._1975} of the AS\&E-MIT group. In particular: the UVT instrument consisted of a Cassegrain telescope followed by a grating spectrometer of the Wadsworth-type; the Utrecht soft X-ray (USXD) consisted of a grazing-incidence parabolic collector while the Utrecht medium X-ray detector (UMXD) was a 1.7 $\umu$ titanium proportional counter; the HXX experimental package contained three major components: a collimator assembly, a large area detector (LAD) unit, and a Bragg-crystal spectrometer (BCS) tuned for detection of the silicon lines. The details of these experiments are summarized in Table \ref{tab:ANS}.
 ANS scientific highlights include the discovery of X-ray bursts, flash of X-rays of several seconds, emitted by neutron stars in binary accreting systems \cite{Heise_et_al._1976}, the detection of X-rays from Stellar Coronae (Capella) \cite{Mewe_et_al._1975}, the first detection of X-ray flares from UV Ceti and YZ CMi \cite{Heise_et_al._1975}
 
\begin{table}[h]%
\caption{ANS \label{tab:ANS}}
\centering
\begin{tabular}{ l c c c c c c } 

\hline

\multirow{2}{*}{Instrument}& UVT&\multicolumn{2}{c}{SXX} &\multicolumn{3}{c}{HXX} \\
&& \textsc{usxd}& \textsc{umxd}&\multicolumn{2}{c}{\textsc{lad}} &\textsc{bcs}\\

\hline
Bandpass (keV) & & $0.2-0.28$ & $1-7$ & \multicolumn{2}{c}{$1-30$}&$1-4.2$ \\
Bandpass (\AA) &1550\AA$-$3300\AA &&&&&\\
Eff Area (cm$^2$) &266&144&45&\multicolumn{2}{c}{40}&6 \\
FOV (FWHM)&2.5$^{\prime}\times$2.5$^{\prime}$&$34^\prime$&38$^{\prime}\times$75$^{\prime}$&10$^{\prime}$&3$^\circ$&3$^\circ$\\

\hline
\end{tabular}
\end{table}

\subsection{\textsc{ariel V}}
The Ariel V Satellite, developed by a joint collaboration of UK and US, was launched from the San Marco platform on October 15, 1974, into a low inclination ($2.8^{\circ}$), near-circular orbit at an altitude of $\sim$ 520 km. The orbital period was 95 min. The mission ended on March 14, 1980. The British Science Research Council managed the project for the UK, The NASA GSFC for the US. Ariel V was dedicated to the monitoring of the X-ray sky. The science payload included 6 instruments. Four, aligned with the spin axis, were devoted to a detailed study of a small region of the sky within $\sim$10$^\circ$ of the satellite pole. The set included: a rotation modulation collimator (RMC or Exp-A), consisting of rotation collimators and three different  detectors, a photo-multiplier, an electron multiplier and a proportional counter; a high resolution proportional counter spectrometer (Exp-C); a Bragg Crystal Spectrometer (Exp-D) operating in the energy band 2-8 keV, that used a honeycomb collimator; and a scintillator telescope (ST or Exp-F). The remaining 2 instruments were arranged in a direction perpendicular to the spin axis. The all-sky monitor (ASM or Exp-G), the only experiment of the mission developed by the US, utilized two X-ray pin-hole cameras to image the sky in order to monitor transient X-ray phenomena, and all the strong X-ray sources for long-term temporal effects; the Sky Survey Instrument (SSI or Exp-B) \cite{Villa_et_al._1976} consisted of two pairs of proportional counters (LE and HE) \cite{Smith_and_Courtier_1976}. 
Ariel V performed long-term monitoring of numerous X-ray sources. It also discovered several long period (minutes) X-ray pulsars \cite{White_et_al._1978} and several bright X-ray transients probably containing a Black Hole (e.g. A0620-00=Nova Mon 1975) \cite{Pound_et_al._1976}, \cite{Elvis_et_al._1975}. It also discovered iron line emission in extragalactic sources \cite{Sanford_et_al._1975} and established Seyfert I galaxies (AGN) as a class of X-ray emitters. In Table \ref{tab:Ariel V} we report details of the scientific payload of the mission.
\begin{table}[h]%
\caption{Ariel V payload consisted of 6 instruments; 4 aligned with the spin axes (Exp A, Exp C, Exp D, Exp F) and two were offset (Exp G, Exp B). \label{tab:Ariel V}}
\centering
\begin{tabular}{ l c c c c c c c} 

\hline

\multirow{3}{*}{Instrument}& \multicolumn{4}{c}{Aligned}&\multicolumn{3}{c}{Offset}   \\
& A \textsc{(rmc)} & C&D &F \textsc{(st)}&G \textsc{(asm)}&\multicolumn{2}{c}{B \textsc{(ssi)}}  \\
&&&&&&\textsc{le}&\textsc{he}\\

\hline
Bandpass (keV) &0.3-30 &1.3-28.6&2-8&26-1200&3-6&1.2-5.8&2.4-19.8  \\

Eff Area (cm$^2$) &--&--&--&8&&\multicolumn{2}{c}{290} \\
FOV (FWHM) &10$^\circ$-20$^\circ$&3.5$^{\circ}$&&\\

\hline
Energy range (keV)&\multicolumn{4}{c}{0.7-10}\\
\hline

\end{tabular}
\end{table}

\subsection{\textsc{Cos-b}}
Cos-B was an ESA mission built by the so called Caravane Collaboration that included: The Laboratory for Space Research, Leiden, The Netherlands; Istituto di Fisica Cosmica e Informatica del CNR, Palermo, Italy; Laboratorio di Fisica Cosmica e Tecnologie Relative del CNR, Milano, Italy; Max-Planck-Institut fur Extraterrestrische Physik, Garching, Germany; Service d'Electronique Physique, CEN de Saclay, France; Space Science Department of ESA, ESTEC, Noordwijk, The Netherlands

The principal scientific objective was to provide a view of the gamma-ray Universe, nevertheless it took onboard  a proportional counter sensitive to 2-12 keV X-rays. As one can read in \cite{Bennet_1990} "This detector was intended to provide synchronization of possible pulsed gamma-ray emission from pulsating X-ray sources. The pulsar synchronizer was also used for monitoring the intensity of radiation from X-ray sources." 



\subsection{\textsc{ariel VI}}
UK6, named Ariel VI after launch, was launched from the Wallops Island Launch Center in the USA on June 2, 1979. The orbit was elliptical with an apogee of 650 km and a perigee of 600 at an inclination of 55 $^\circ$. Ariel VI was a national UK mission but, in comparison with the success of its predecessor Ariel V, much less successful due to the problems caused by the interference with powerful military radars. In fact strong magnetic fields severely hammered the command encoder and the pointing operations. Ariel VI carried 3 scientific instruments: one was a cosmic ray experiment consisting of Cerenkov and gas scintillation counters, and the other 2 were X-ray instruments. The soft X-ray Telescope (here SXT) developed by MSSL in collaboration with the University of Birmingham, consisted of four grazing-incidence hyperboloid mirrors that reflected X-rays through an aperture/filter to four continuous-flow propane gas detectors \cite{Cole_et_al._1981}. The medium X-ray proportional counter (MXPC) developed by the Leicester University consisted of 4 multi-layered Xe-proportional counters \cite{Hall_et_al._1981}. Ariel VI continued to observe until February 1982. Table \ref{tab:Ariel VI} shows some of the features of the X-ray instruments. Although partially, the observations carried with Ariel VI brought some results like the observation of phase variable iron line emission of the source GX 1+4 \cite{Ricketts_et_al._1982} or the spectral observation of Active Galaxies \cite{Hall_et_al._1981}

\begin{table}[h]%
\caption{Ariel VI featured two X-ray instruments onboard the SXT and the MXPC. \label{tab:Ariel VI}}
\centering
\begin{tabular}{ l c c } 

\hline

\multirow{2}{*}{Experiment}& SXT&\multirow{2}{*}{MXPC}  \\
& Grazing telescope + Xe-prop. counter &   \\
\hline
Bandpass (keV) &0.1--2 &1--50 \\

Eff Area (cm$^2$) &65&300\\

FOV (FWHM) &1.2$^{\circ}$--4.6$^{\circ}$&3$^{\circ}$\\

\hline
\end{tabular}
\end{table}

\subsection{\textsc{exosat}}
  
The European Space Agency's (ESA's) X-ray Observatory, EXOSAT \cite{Taylor_et_al._1981}, was active from May 1983 to April 1986. It was launched into a highly eccentric orbit (e$\sim$ 0.93) with a 90.6 hr period, inclination of 73$^\circ$, at an apogee and perigee of 191,000\,km and 350\,km respectively (at the beginning of the mission). This - at that time - peculiar orbit was chosen to enable long (from 76 hr or 90 hr), uninterrupted observations during a single orbit. Due also to the great distance from the Earth ($\sim$50 000\,km), EXOSAT was almost always visible from the ground station at Villafranca in Spain during science instruments operations.
The payload of the satellite consisted of the Low Energy Telescopes (LE), composed of two identical Wolter-I telescopes. Each could operate in imaging mode by means of Channel Multiplier Array (CMA) or Position Sensitive Detector (PSD) or in spectroscopy mode with gratings behind the optics and the CMA in the focal plane, \cite{DeKorte_et_al._1981}; the Medium Energy Instrument (ME), the main instrument in the lunar occultation mode  \cite{Turner_Smith_and_Zimmermann_1981}; the Gas Scintillation Proportional Counter (GSPC \cite{Peacock_et_al._1981}). The characteristics of the instruments are shown in Table \ref{tab:exosat}.
During the performance verification phase the PSDs of the two LE failed. About half a year later one of the channel plates failed. However, overall the LE functioned up to the end, and discoveries were made using the X-ray grating spectrometers (built by SRON)\cite{Bleeker_and_Verbunt_2013,Bleeker_2022}. 
Most notable were the discovery of quasi-periodic oscillations (QPOs) of low mass X-ray binaries, the soft excesses from AGN, the red and blue shifted iron K line from SS433, the characterization of many orbital periods of low mass X-ray binaries, and the discovery of new transient sources. 

The scientific highlights of the EXOSAT  mission are reported in a special issue of the Memorie della Società Astronomica Italiana \cite{MSAI_1988}.
\begin{table}[h]%
\caption{EXOSAT's science payload: the LE, ME and GSPC. \label{tab:exosat}}
\centering
\begin{footnotesize}
\begin{tabular}{ l c c c c c } 

\hline

\multirow{2}{*}{Instrument}& \multicolumn{3}{c}{LE}&\multirow{2}{*}{ME}&\multirow{2}{*}{GSPC}  \\
& \textsc{cam} & \textsc{psd}&Spectrometer &&  \\
\hline
Bandpass (keV) &0.04--2 &0.1--2&--&1--50&2--20  \\

Eff Area (cm$^2$) &0.4--10&--&--&1800&$\sim$ 10--100\\
FOV &2.2$^\circ$&1.5$^\circ$\\
\multirow{2}{*}{Angular res (FWHM)}&\multicolumn{2}{c}{on axis}&\multirow{2}{*}{--}&\multirow{2}{*}{45$^\prime\times45^\prime$}&\multirow{2}{*}{45$^\prime$}\\
&12$^{\prime\prime}$&50$^{\prime\prime}$&&&\\
\multirow{2}{*}{Energy res (FWHM)}&\multirow{2}{*}{none}&\multirow{2}{*}{$\Delta$E/E= 44/E (keV)$^{1/2}$ \% }&&21\% at 6 keV (Ar)&\multirow{2}{*}{27/E (keV)$^{1/2}$ \%  }\\
&&&&18\% at 22 keV (Xe)&\\ 

\hline

\end{tabular}
\end{footnotesize}
\end{table}

\section{Late 1970s and the 1980s: The program in Japan}

Japan, thanks to the leadership of Minoru Oda, contributed to X-ray astronomy with several missions, becoming a well recognized country in space science. The first of those missions was CoRSa-b renamed, after the successful launch, Hakucho. 

\subsection{\textsc{hakucho}}

Hakucho, Japanese for swan (like one of the archetypal X-ray sources, Cyg X-1) developed by the Institute of Space and Astronautical Science (ISAS), was launched from the Kagoshima Space Center on February 21, 1979. It was placed into a near-circular orbit with an apogee of 572 km, a perigee of 545 km, an inclination of $29.9^{\circ}$ and an orbital period of 96 minutes \cite{Oda_1980}. It was the second of the series CoRSa (Cosmic Radiation Satellite)\footnote{Unfortunately the first of the satellite of the series Corsa-a failed to reach the orbit.}. The mission ended on April 16, 1985. Its main goal was the study of transient phenomena using three different instruments: the Very Soft X-ray experiment (VSX), based on four units of proportional counters with very thin polypropylene windows. Two of the counters were oriented along the spin axis (VXP) and two were offset (VXV); the Soft X-ray instrument (SFX) that included six proportional counters with Beryllium windows. Two were equipped with a coarse modulation collimator (CMC), two with a fine modulation collimator (FMC), and the last two aimed at scanning the sky (SVC) operated in offset mode; the Hard X-ray (HDX) detector that consisted of 2 Na(T1) scintillators with an offset of 2.7$^\circ$.  
   Table \ref{tab:hakucho} summarizes the principal characteristics of the mission payload \cite{Inoue_et_al._1980}. Hakucho data led to the discovery of many burst sources, and the soft X-ray transient sources Cen X-4 and Apl X-1.
    \begin{table}[h]%
\caption{HAKUCHO (CoRSa-b). The payload carried three instruments for the detection of very soft (VSX), soft (SFX), and hard (HDX) X-rays. \label{tab:hakucho}}
\centering
\begin{tabular}{ l c c c c c c} 
\hline
\multirow{2}{*}{Instrument} &  \multicolumn{2}{c}{VSX} & \multicolumn{3}{c}{SFX}  &HDX  \\
& \textsc{vxp}& \textsc{vxv}&\textsc{cmc}&\textsc{fmc}&\textsc{svc}&\\
\hline
Bandpass (keV) &\multicolumn{2}{c}{0.1--1}&\multicolumn{3}{c}{1.5--30}&10--100\\

\multirow{2}{*}{Eff Area (cm$^2$)}&\multicolumn{2}{c}{\multirow{2}{*}{$\sim$ 78}}  & \multirow{2}{*}{69}&\textsc{fmc}1 40&\multirow{2}{*}{32} & \multirow{2}{*}{$\sim$ 45}\\
      & &&& \textsc{fmc}2 83&& \\

FOV (FWHM) &  $ 6.3 ^{\circ} \times 2.9 ^{\circ} $ &  $17.6 ^{\circ} $ &
      $ 24.9 ^{\circ} \times 2.9 ^{\circ} $   & $ 5.8 ^{\circ} $ & $ 4.4 ^{\circ} \times 10.0 ^{\circ} $ 
     & $ 50.3 ^{\circ} \times 1.7 ^{\circ} $  \\
\hline     
\end{tabular}
\end{table}
   
\subsection{\textsc{Hinotori}} Hinotori, Japanese for Phoenix or Firebird, was the first of the series of Astra satellites. It was dedicated to the study of solar phenomena, in particular to solar flares during the solar maximum. It was launched from the Kagoshima Space Center (now Ichinoura) on February 21, 1981, and operated until October 8, 1982. The orbit was near-circular with an apogee altitude of 603\,km, perigee of 548\,km, inclination of 31.3$^{\circ}$, and period of 96.20 min. For the solar flare studies, Hinitori carried onboard the Solar X-ray Telescope (SXT), equipped with two sets of bi-grid modulation collimators for the imaging of the hard X-ray emission, using the rotating modulation collimator technique. In addition, the Solar X-ray Aspect Sensor (SXA) was a system of collimating lenses to determine the flare position with a resolution of 5 arcsec.  The Soft X-ray Crystal Spectrometer (SOX, \cite{Tanaka_et_al._1982}) enabled the spectroscopy of X-ray emission lines from highly ionized iron during a flare. It consisted of coarse (SOX1) and fine (SOX2) Bragg spectrometers. Three additional instruments enabled the monitoring of the flares over a large energy band: the Soft X-ray Flare Monitor (FLM), the Hard X-ray Flare Monitor (HXM), and the Solar Gamma Ray Detector (SGR \cite{Yoshimori_et_al._1983}). The FLM was a gas (Xe) scintillation proportional counter while the HXM and SGR detectors were NaI(T1) and CsI(T1) scintillation counters, respectively. The 8 counters of the HXM instrument had different characteristics as reported in Table \ref{tab:Hinotori} \cite{TanakaYasuo_1983}.  In addition to the aforementioned instruments, Hinotori hosted a particle ray monitor (PXM), a plasma electron density measurement instrument (IMP), and a plasma electron temperature measurement instrument (TEL). All instruments were co-aligned with the spacecraft spin (Z) axis that was set 1$^\circ$ off the sun center and therefore no additional driving mechanism for the detectors was necessary. The main scientific results of Hinotori include: the time profile and spectrum of the X-ray flares \cite{Tanaka_1987}, \cite{Yoshimori_1990}; monitoring of the electron flux above 100 keV; discovery of high-temperature phenomena reaching up to 50 million $^\circ$C and clouds of light-speed electrons floating in coronas \cite{Oyama_Schlegel_and_Watanabe_1988}. 

    \begin{table}[h]%
\caption{HINOTORI (ASTRO-A). The payload consisted of a solar flare telescope (SXT), an analyzer (SXA), a spectrometer for soft X-ray (SOX), monitors for hard X-rays (HXM) and gamma-rays (SGR), for both solar flares (FLM) and for particle emission PXM).  \label{tab:Hinotori}}
\centering
\begin{tabular}{l c c c c c c c c c} 
\hline

\multirow{2}{*}{Instrument} & SXT  &SXA & \multicolumn{2}{c}{SOX}& FLM& \multicolumn{2}{c}{HXM}  &SGR &PXM \\ 
&&&\textsc{sox1}&\textsc{sox2}&&\textsc{hxm1}&\textsc{hxm2}&&\\
 \hline
Bandpass (keV)&17--40&&1.72-1.99{\AA}&1.83-1.89{\AA}&2--12&17--40&40--340&200--6 700 &100-800\\
Area (cm$^2$)&113& &6.69{\AA} &2.36{\AA}&0.5&\multicolumn{2}{c}{57}&62&2.2  \\
Ang. Res. (FWHM)&30$^{\prime\prime}$&5$^{\prime\prime}$&&&&
\\
Time Res (ms)&$\sim$ 6$\times$10$^3$&&\multicolumn{2}{c}{6--10$\times$10$^3$}&125&7.8&125&128ch/4s&125/ch\\
Energy Res &&&2m{\AA}&0.15m{\AA}&&&&0.1 E$^{1/2}$ MeV&\\




\hline
\end{tabular}
\end{table}
    \subsection{\textsc{tenma}} 
    Tenma, Japanese for Pegasus, developed by ISAS, was the second satellite of the ASTRO series (ASTRO-B), and the second Japanese satellite for X-ray astronomy. It was launched on February 1983 and placed into a near-circular orbit with apogee of 501 km, perigee of 497 km, and inclination angle of 31.5 degrees. The orbital period was 96 minutes. Its scientific payload consisted of four instruments: a scintillation proportional counter (SPC), an X-ray focusing collector (XFC), a transient sources monitor (TSM), and a radiation belt and Gamma-ray monitor (RBM/GBD,  \cite{Tanaka_et_al._1984}). In particular, the SPC, devoted to spectral and temporal studies, consisted of 10 GSPC divided in 3 groups (SPC-A, B, C) of four, four, and two units, respectively. The XFC, consisting of mirrors and position sensitive proportional counters, was designed to observe very soft X-ray sources. The TSM served as an X-ray monitor because of its wide FOV. It included two detector groups: an Hadamard X-ray telescope system (HXT) and a scanner counting system (ZYT). Two small scintillation counters monitored the non-X-ray background and the gamma-ray burst emissions. The entire payload was mostly aligned with the stabilized spin-axes (Z) of the satellite. Detailed information about the instruments are reported in Table \ref{tab:Tenma}. Tenma observations continued intermittently until 11 November 1985. The main results of the mission was the discovery and study of the iron line region of many classes of sources. 

Tenma science highlights include: The Discovery of hot plasma of several tens of millions of degrees located along the Galactic plane \cite{Koyama_1989}; The discovery of the iron absorption line in the energy spectra of X-ray bursts, which was red-shifted in the strong gravitational field of the neutron star \cite{Waki_et_al._1984}, \cite{Suzuki_et_al._1984}, \cite{Inoue_1985}; The identification in low-mass X-ray binaries of X-ray emission regions on the surface of the neutron star and in the accretion disk \cite{Mitsuda_et_al._1984}.
    \begin{table}[h]%
\caption{Tenma.The payload consisted of four instruments: the SPC, the XFC, the TSM, and a the RBM/GBD monitor. \label{tab:Tenma}}
\centering
\begin{tabular}{l c c c c c c c} 
\hline
\multirow{2}{*}{Instrument} & \multicolumn{3}{c}{SPC}  &  XFC   &\multicolumn{2}{c}{TSM} & RBM/GBD \\
&\textsc{spc-a}&\textsc{spc-b}&\textsc{spc-c}&&\textsc{hxt}&\textsc{zyt}&\\
\hline
Bandpass (keV) &\multicolumn{3}{c}{2-60}&0.1-2  &2-25 &1.5-25&10-100 \\
Energy Res (FWHM)&\multicolumn{3}{c}{9.5\% at 5.9 keV}&&&\\
Area (cm$^2$)&\multicolumn{2}{c}{320}&80&15&114&280&14\\

(FOV) FWHM&3.1$^\circ$ &2.5$^\circ$&3.8$^\circ$&1.4$^\circ\times$5$^\circ$& 40$^\circ\times$40$^\circ$&2$^\circ\times$25$^\circ$& 1sr\\
 
 \hline
\end{tabular}
\end{table}

    \subsection{\textsc{Ginga}} 
    Ginga, Japanese for Galaxy, ASTRO-C before launch, was launched on Feb 5, 1987 and operated until November 1, 1991. Astro-C was the result of a collaboration between Japanese research institutions, the University of Leicester and the Rutherford-Appleton Laboratory in the UK, and the Los Alamos National Laboratory (USA). Ginga followed a near circular orbit at a perigee distance of 505 km and an apogee of 675 km. It was originally planned to make a circular orbit of 630 km but atmospheric conditions at launch constrained the satellite into an elliptic orbit. 
    The inclination of the orbit was 31$^\circ$, and the period was 96 min. The primary mission objective was the study of the time variability of X-rays from active galaxies such as Seyfert galaxies, BL Lac objects, and quasars in the energy range 1.5-30 KeV. Accurate timing analysis of galactic X-ray sources was also one of the goals of the mission \cite{Makino_and_Astro-C_Team_1987}. The payload of the satellite consisted of three instruments: a Large Area Proportional Counter (LAC, \cite{Turner_et_al._1989}), an All Sky Monitor (ASM, \cite{Tsunemi_et_al._1989}), and a Gamma-ray Burst Detector (GBD, \cite{Murakami_et_al._1989}). The LAC consisted of eight multi-cells proportional counters. The ASM consisted of 2 identical gas proportional counters. Each counter was equipped with a collimator which had 3 different FOVs. The GBD included two detectors: a proportional counter and a scintillation spectrometer. The characteristics of these instruments are summarized in Table \ref{tab:Ginga}.
   
 \begin{table}[h]%
\caption{Ginga (ASTRO-C). The primary instrument on-board was the LAC. The ASM and the GBD completed the payload. \label{tab:Ginga}}
\centering
\begin{tabular}{l c c c c} 
\hline

  \multirow{2}{*}{Instrument} &  LAC  &ASM& \multicolumn{2}{c}{GBD} \\
&&&\textsc{pc}&\textsc{sc}\\
\hline
Bandpass (keV) & 1.5-37  &1-20&\multicolumn{2}{c}{1.5-500}  \\
Energy Res (FWHM)&18\% at 6 keV&&\\
Area (cm$^2$)&4500&70&63&60\\
FOV (FWHM)&$ 0.8 ^{\circ} \times 1.7 ^{\circ} $&$ 1 ^{\circ} \times 45 ^{\circ} $&\multicolumn{2}{c}{all sky}\\
 
 \hline
\end{tabular}
\end{table}
\section{Late 1970s and the 1980s: The program in Russia and India}

The X-ray astronomy program of the Soviet Union had modest beginnings in the 1970s with the FILIN X-ray experiment aboard the Salyut-4 space station. It continued in the 1980s with experiments onboard the Astron (1983-1988) and Mir (1987-2000) space stations. This latter program had a strong European involvement. These programs generally suffered from a limited
observation time allocation because of other commitments of the manned spacecraft. During the late 1970s and into the 1980s, the Soviet program was focused on studies of gamma-ray bursts. The Konus experiments on the Venera 11-14
spacecraft yielded major advances in this field \cite{Mazets_1981, Higdon_Lingenfelter_1990}. A notable result at X-ray wavelengths was the discovery of an unusual gamma burst on March 5, 1979 with sustained X-ray emission that exhibited periodic pulsations \cite{Mazets_1979}. 
\subsection{\textsc{filin/salyut-4}}

 The FILIN X-ray instrument aboard the manned orbiting space station Salyut-4 (December 1974) consisted of three detectors sensitive in the 2-10~keV range \cite{Babichenko_1977} and a smaller proportional counter for soft X-ray studies (0.2-2~keV), with a rather large FOV (see Table \ref{tab:filin}).
 
 Gas flow proportional counters were used as the detectors. A gas flow system supplied a gas mixture for the counters. To determine the source coordinates, two star sensors were installed. The X-ray detectors, all optical sensors, and the gas flow system were mounted on the outside of the station, while the power supply and electronics was inside. 
 Scanning observations were carried out for about 1 month and pointed observations for about 2 months; studies included observations of Sco X-1, Her X-1, and Cyg X-1 \cite{Babichenko_1977} and the X-ray nova A0620-00 \cite{Bradt_1992,Martynov_et_al._1975} 
 \begin{table}[h]%
\caption{FILIN. The x-ray instrument on-board the Salyut-4  space station. \label{tab:filin}}
\centering
\begin{tabular}{p{4.0cm} p{3.cm} p{3.cm}} 
\hline

Instrument &  sFilin  & Filin  \\

\hline
Bandpass (keV) & 0.2-2  &2-10  \\
Area (cm$^2$)&40&450\\
FoV (FWHM)&\multicolumn{2}{c}{\hspace{-2cm}$3^{\circ}\times10^{\circ}$}\\
 \hline
\end{tabular}
\end{table}
 
\subsection{\textsc{skr-02m}} 
The experiment SKR-02M on the Astron station (1983)
consisted of a large proportional counter of effective area $\sim0.17$\,m$^2$) sensitive from 2 to 25\,keV \cite{Babichenko_1990}. The field of view was 3$^{\circ}\times$ 3$^{\circ}$ (FWHM). Data were sent via telemetry in 10 energy channels. Results have been
reported from studies of the Crab nebula and pulsar, Her X-1, A0535 + 26,
and Cen X-3 \cite{Babichenko_1990}. The prolonged low state of Her X-1
in 1983 was studied, and the 1984 turn-on was reported \cite{Giovanelli_1984}.


\subsection{\textsc{xvantimir}} 
The R\"ontgen X-ray observatory was launched in 1987 aboard the Kvant module which docked to the MIR space station. The complement of detectors \cite{Sunyaev_1990} included a sensitive high energy 15-200\,keV X-ray experiment (HEXE, \cite{Reppin_1985}), a coded-mask system for imaging high-energy photons (TTM, \cite{Brinkman_1985}), and a gas scintillation proportional counters (GSPC, \cite{Smith_1985}). It also carried two gamma-ray experiments, which reached down to 30-40\,keV \cite{Sunyaev_1990}. Röntgen was an international endeavor with contributions from Germany, UK, ESA, and the Netherlands. The highlight of the mission was the discovery and study (with Ginga) of the X-ray emission from SN1987A \cite{Sunyaev_1987}. A high energy tail in the spectrum of the X-ray nova GS2000 + 25 was discovered \cite{Sunyaev_1988a} further indicating its similarity to the black-hole candidate A0620-00. Timing results for the Her X-1 pulsar in 1987-1988 showed it to be continuing its spin up \cite{Sunyaev_1988b}. The principal characteristics of the R\"ontgen X-ray observatory are reported in Table \ref{tab:Roentgen}.
 \begin{table}[h]%
\caption{R\"ontgen X-ray observatory.  \label{tab:Roentgen}}
\centering
\begin{tabular}{p{4.0cm} p{2.cm} p{2.cm} p{2.cm}} 
\hline
Instrument &  TTM &GSPC& HEXE \\
\hline
Energy range (keV) & 2-32  &4-100&15-200 \\
Energy resolution (FWHM) & 18\% at 6~keV &10.5\% at 6~keV& 30\% at 60~keV\\
Area (cm$^2$)&600&300&800\\
FOV (FWHM) &$7.8 ^{\circ}\times7.8^{\circ}$&$3^{\circ}\times3^{\circ}$ &$1.6^{\circ}\times1.6^{\circ}$\\
 \hline
\end{tabular}
\end{table}

\subsection{\textsc{aryabhata}}
Aryabata, named after the Indian mathematician and astronomer of the fifth century, was the first satellite of India completely designed and built by the Indian Space Research Organization (ISRO). It was launched on April 19, 1975 from the Russian rocket launch site Kaputsin Yar. Its orbit had a perigeee of 563\,km, an apogee of 619\,km, and inclination of 50.7$^{\circ}$. The period was of 96.46\,min. The mission ended on March 1981 and the satellite reentered the Earth's atmosphere on February 10, 1992. Three instruments dedicated to: Aeronomy, solar physics, and X-ray Astronomy were on-board. The X-ray detector consisted of a proportional counter filled with a mixture of Ar, CO$_{2}$ and He, and operated in parallel mode, in the energy range from 2.5 to 115 \,keV. The effective area was $\sim 15.4$\,cm$^2$ and the FOV, circular, with $12.5^{\circ}$ (FWHM). In particular Aryabhata made observation of Cyg X-1, finding a hardening in its spectrum, \cite{Rao_et_al._1976}, and of other two X-ray sources, namely GX17+2, and GX9+9 \cite{Kasturirangan_et_al._1976}. 


\subsection{\textsc{bhaskara}}
Two satellites Bhaskara I and II were developed by ISRO and named after the two famous Indian mathematicians Bhaskara (or Baskara I) of the 7th century, and Bhaskara II (or Bhaskaracharya, Bhaskara the teacher) of the 12th century. We will report here only about Bhaskara I, since Baskara II didn't carry X-ray instruments. Bhaskara I, was launched on June 7, 1979 from Kaputsin Yar. Its orbital perigee and apogee were 512\,km and 557\,km respectively, the inclination was of 50.7$^{\circ}$ and period 95.20\,min. The mission ended on February 17, 1989, after almost 10 years. The main objectives of the mission were: 1) to conduct observations of the earth yielding data for hydrology, forestry, and geology applications; 2) to conduct ocean-surface study using a SAtellite MIcrowave Radiometer (SAMIR), and 3) among other minor investigations, to conduct investigation in X-ray astronomy. The X-ray instrument consisted by a Pinhole X-ray Survey camera operating in the energy range between 2 and 10 keV with the purpose of observing transient sources and long term variability of steady sources. At the image plane of the camera there was a position-sensitive proportional counter, the detector operated with success during  the first month after launch. However, it had to be turned off, and when, after some time, was turned on again it didn't operate correctly; the reason of the malfunction was never found.

\section{The golden age of X-Ray astronomy, from the 1990s to the present}

The 1990s can be considered as a sort of renaissance of X astronomy. It consisted of years of significant missions, that brought X astronomy into its full maturity. 
The decade begins with the launch of the Soviet mission Granat (December 1989) and of the German mission ROSAT (June 1990), and soon after the Japanese ASCA. In the mid-1990s, BeppoSAX and RXTE were launched and in the late 1990s, Chandra and XMM-Newton. 
 \subsection{The program in the US}
 \subsection{\textsc{ulysses}}

The Ulysses mission was a joint mission between NASA and ESA to explore the solar environment at high ecliptic latitudes. Launched on October 6, 1990, it reached Jupiter for its 'gravitational slingshot' in February 1992. It passed the south solar pole in June 1994 and crossed the ecliptic equator in February 1995. In addition to its solar environment instruments, Ulysses also carried onboard plasma instruments to study the interstellar and Jovian regions, as well as two instruments for studying X- and Gamma-rays of both solar and cosmic origins. The mission could send data in 4 different telemetry modes at rates of 128, 256, 512, and 1024 b/s. The time resolution of the Gamma-ray burst instrument depended on the chosen data rate. The maximum telemetry allocation for the instrument was about 40 b/s.

The Ulysses solar X-ray and cosmic gamma-ray burst experiment (GRB) had 3 main objectives: study and monitor solar flares, detect and localize cosmic gamma-ray bursts, and in-situ detection of Jovian auroras. Ulysses was the first satellite carrying a gamma burst detector which went outside the orbit of Mars. This resulted in a triangulation baseline of unprecedented length, thus allowing major improvements in burst localization accuracy. The instrument was turned on 9 November 1990. The GRB consisted of 2 CsI scintillators (called the Hard X-ray detectors)and 2 Si surface barrier detectors (called the Soft X-ray detectors). The detectors were mounted on a 3\,m boom to reduce background generated by the spacecraft's radioisotope thermoelectric generator.
The hard X-ray detectors operated in the range 15-150\,keV. The detectors consisted of two 3 mm thick by 51 mm diameter CsI(Tl) crystals mounted via plastic light tubes to photomultipliers. The hard detector varied its operating mode depending on the measured count-rate, the ground command, or a change in spacecraft telemetry mode. The trigger level was normally set for 8-sigma above background corresponding to a sensitivity 2$\times 10^{-6}$ erg cm$^{-2}$ s$^{-1}$ \cite{Boer_Sommer_Hurley_1990}. When a burst trigger was recorded, the instrument switched to high resolution data, recording a 32-kbit memory for a slow telemetry read out. Burst data consisted of either 16 s of 8-ms resolution count rates or 64 s of 32-ms count rates from the sum of the 2 detectors. There were also 16 channel energy spectra from the sum of the 2 detectors (taken either in 1, 2, 4, 16, or 32 second integration). During 'wait' mode, the data were taken either in 0.25 or 0.5 s integration and 4 energy channels (with shortest integration time being 8 s). Again, the outputs of the 2 detectors were summed.
The soft X-ray detectors consisted of two 500 $\mu$m thick, 0.5 cm$^{2}$ area Si surface barrier detectors. A 100 mg cm$^{-2}$ beryllium foil front window rejected the low energy X-rays and defined a conical field of view of 75$^{\circ}$ (half-angle). These detectors were passively cooled and operated in the temperature range $-35$ $^{\circ}$C to $-55$  $^{\circ}$C. This detector had 6 energy channels, covering the range 5-20 keV. Ulysses results have been mainly about the Sun and its influence on nearby space \cite{Marsden_and_Angold_2008}. 

 \subsection{\textsc{bbxrt}}

The Broad Band X-ray Telescope (BBXRT, \cite{Serlemitsos_et_al_1984}) was flown on the space shuttle Columbia (STS-35) as part of the ASTRO-1 payload (December 2, 1990 - December 11, 1990). It was designed and built by the Laboratory for High Energy Astrophysics at NASA/GSFC. 
BBXRT was the first focusing X-ray telescope operating over a broad energy range 0.3-12\,keV, with moderate energy resolution (90\,eV at 1\,keV and 150\,eV at 6\,keV). It consisted of two identical co-aligned telescopes each with a segmented Si(Li) solid state spectrometer (detector A and B) with five pixels. The telescope consisted of two sets of nested grazing-incidence mirrors, whose geometry was close to the ideal paraboloidal/ hyperboloidal surfaces (modified Wolter type-I ). This simplified fabrication and made possible nesting many shells to yield a large geometric area. The effective on-axis areas was 0.03\,m$^2$ at 1.5\,keV and 0.015 m$^2$ at 7\,keV. The focal plane consisted of a 5-element lithium drifted silicon detector with energy resolution of about 100\,eV FWHM. 
Despite operational difficulties with the pointing systems, the BBXRT obtained
high quality spectra from some 50 selected objects \cite{Serlemitsos_et_al_1992}, both Galactic and extragalactic. 
Results included the resolved iron K line in the binaries Cen X-3 and Cyg X-2 \cite{Smale_et_al_1993}, 
evidence of line broadening in NGC 4151 \cite{Weaver_et_al_1992}, and the study of cooling flow in clusters \cite{Arnaud_et_al_1998}. Details are reported in Table \ref{tab:bbxrt}.
\begin{table}[h]%
\caption{BBXRT\label{tab:bbxrt}}
\centering
\begin{tabular}{ p{4.5cm} p{3.5cm} p{3.5cm}} 
\hline

\hline
Instrument & BBXRT on STS-35  \\
\hline
Bandpass~(keV) &0.3-12  \\

Eff Area~(cm$^2$) (at 1.5 keV)&765  \\
Eff Area~(cm$^2$) (at 7 keV)  &300 \\
FOV (diameter)& 17.4$^{\prime}$ \\
Central pixel FOV diameter& 4$^{\prime}$ \\
Angolar resolution &$2^{\prime}-6^{\prime}$\\
Energy resolution (eV, FWHM) at 1 keV &90\\
Energy resolution (eV, FWHM) at 6 keV &150\\
\hline
\end{tabular}
\end{table}
 

 \subsection{\textsc{Rxte}}
 
 The Rossi X-ray Timing Explorer (RXTE, \cite{Bradt_et_al_1993}) was launched on December 30, 1995 from the NASA Kennedy Space Center. The mission was managed and controlled by NASA/GSFC. RXTE featured unprecedented time resolution in combination with moderate spectral resolution to explore the time variability of the X-ray sources. Time scales from microseconds to months were studied in the spectral range from 2 to 250\,keV. Originally designed for a required lifetime of two years with a goal of five, RXTE completed 16 years of observations (!) before being decommissioned on January 5, 2012.

The spacecraft was designed and built by the Applied Engineering and Technology Directorate at NASA/GSFC. The launch vehicle was a Delta II rocket that put RXTE into a low-earth circular orbit at an altitude of 580\,km, corresponding to an orbital period of about 90 minutes, with an inclination of 23 degrees.
Operations were managed at GSFC. 


The mission carried onboard two pointed, collimated instruments: the Proportional Counter Array (PCA, \cite{Zhang_et_al_1993}) developed by GSFC to cover the lower part of the energy range, and the High Energy X-ray Timing Experiment (HEXTE, \cite{Gruber_et_al_1996}) developed by the University of California at San Diego, covering the upper energy range. 
The PCA was an array of five proportional counters with a total collecting area of 6 500\,cm$^2$. Each unit consisted of  a layer Propane veto,  3 layers Xenon, each split in two, and a Xenon veto layer.  

HEXTE consisted of two clusters each containing four NaI/CsI phoswich scintillation counters. Each cluster could 'rock'
along mutually orthogonal directions to provide background measurements (1.5 or 3.0 degrees away from the source) every 16 to 128\,s. Automatic gain control was provided by using a 241Am radioactive source mounted in each detector's field of view. 
Part of the RXTE scientific payload, was an All-Sky Monitor (ASM) from MIT that scanned about 80$\%$ of the sky every orbit, allowing monitoring at time scales of 90 minutes or longer. The ASM~\cite{Levine_et_al_1996} consisted of three wide-angle shadow cameras equipped with proportional counters with a total collecting area of 90\,cm$^2$. The main details of the mission are reported in Table \ref{tab:rxte}.

\begin{table}[h]%
\caption{Rossi XTE\label{tab:rxte}}
\centering
\begin{tabular}{ p{3.5cm} p{2.cm} p{2.cm} p{2.5cm}} 
\hline

\hline
Instrument & ASM  & PCA & HEXTE \\
\hline
Bandpass~(keV)&2-10 &2-60 &15-250\\
Eff Area~(cm$^2$) &90&6 500&2$\times$800 \\
FOV & $6^\circ\times90^\circ$ e.u. &1$^\circ$&1$^\circ$\\

Time resolution &80$\%$ of~the~sky in~90~minutes & 1 $\umu$sec&8 $\umu$sec\\
Energy resolution &&$<$ 18$\%$(6~keV)&15$\%$(60~keV)\\
Spatial resolution &$3^\prime\times15^\prime$& &\\
\hline
Sensitivity (milliCrab)&30 &0.1 & 1000(360~cts)/cluster  \\
Background & &2 mCrab & 50 count/s/cluster   \\
\hline
\end{tabular}
\end{table}
RossiXTE was an extremely successful and productive mission. Science highlights include: The discovery of kilohertz Quasi-Periodic Oscillations (KHz QPOs) in NS systems \cite{klis_et_al._1996}and High Frequency QPOs in BH systems\cite{Morgan_Remillard_Greiner_1997}; The discovery of the first accreting millisecond pulsar, SAX J1808.4–3658\cite{Wijnands_klis_1998}, followed by several more accreting millisecond pulsars; The detection of X-ray afterglows from Gamma Ray Bursts\cite{Bradt_et_al._2003}; the observation of the Bursting Pulsar over a broad range of luminosity, providing stringent test of accretion theories.

 \subsection{\textsc{usa on board argos}}

The Unconventional Stellar Aspect (USA \cite{Wood_2000}) experiment was a multi-purpose experiment based on an X-ray sensor. The main objectives included both basic research in X-ray astronomy and the test of X-ray sensors in space. The experiment was launched on February 23, 1999 from Vandenberg AFB, CA aboard the Advanced Research and Global Observation Satellite (ARGOS). USA operated from April 1999 to November 2000.  It consisted of a pair of large-area gas scintillation proportional counters sensitive to 1–15\,keV mounted on a two-axis pointing system (see Table \ref{tab:USA}). It was a reflight of the SPARTAN 1 instrument flown on the Space Shuttle Discovery in June 1985. USA included precise (roughly microsecond accuracy) time-tagging of events using an integrated GPS receiver. 

 The experiment was used to provide a new atmospheric diagnostic based on atmospheric column density determinations. Energy-resolved photon extinction curves of X-ray celestial sources occulted by the Earth's atmosphere were used. 
 This research was therefore the first to study the neutral atmosphere using X-ray source occultations, and complements UV airglow remote sensing techniques used aboard ARGOS that were insensitive to nighttime neutral density.

\begin{table}[h]%
\caption{USA on board ARGOS \label{tab:USA}}
\centering
\begin{tabular}{ p{4.5cm} p{5.5cm} } 
\hline

\hline
Instrument & USA   \\
\hline
Bandpass~(keV) &0.5-25  \\
Eff Area~(cm$^2$)&2000 at 3 keV  \\
Field of view (FOV) & $1.2^{\circ}\times1.2^{\circ}$ \\
Time resolution $\umu$s& 2 \\
Energy resolution  &0.17 (1 keV at 5.9 keV), 128 raw PHA channel\\
\hline
\end{tabular}
\end{table}

The great majority  of the observations made with USA had either neutron star or black hole sources as targets. Four prominent  transients  active  during  the  life  of  USA  were  XTE~111 18+480,  XTE~Jl550-564,  XTE~Jl859+226  and
GRS~1915+105. . 
\subsection{The program in Europe}
\subsection{\textsc{rosat}}
The Roentgen Satellite, ROSAT, a collaboration between Germany, the US, and the UK, was launched on June 1, 1990 into a near-circular orbit with perigee of 539\,km and apogee 554\,km. The inclination was of 53$^\circ$. The mission operated for almost 9 years until February 12, 1999. The first 6 months of the mission were dedicated to the all sky-survey, followed by the pointing phase. ROSAT obtained the first X-ray and XUV all-sky surveys using an imaging telescope with an X-ray sensitivity of about a factor of 1 000 better than that of UHURU.

The main instrument of the ROSAT observatory was a fourfold nested Wolter-I X-ray telescope (XRT) whose focal plane assembly consisted of a carousel carrying three imaging X-ray detectors. Two of them were position sensitive proportional counters (PSPCs). The third imaging detector was a high resolution imager (HRI) provided by NASA \cite{Truemper_1982}. 
The Wide Field Camera (WFC) was an EUV telescope (with Wolter-Schwarzschild mirrors) designed and built by the UK by the University of Leicester. The focal plane instrumentation of the WFC consisted of a curved micro-channel plate (MCP) \cite{Pounds_and_Wells_1991}. The main information on ROSAT's scientific payload is reported in Table \ref{tab:ROSAT}. 
 \begin{table}[h]%
\caption{ROSAT carried onboard the XRT and a WFC.  \label{tab:ROSAT}}
\centering
\begin{tabular}{l c c c} 
\hline
  Instrument &  \multicolumn{2}{c}{XRT}& WFC \\
Telescope&\multicolumn{2}{c}{Wolter-I}&Wolter-Schwarzschild\\
Focal length (m)&\multicolumn{2}{c}{2.40}&0.525\\
\hline
Detector&PSPC&HRI&MCP\\
\hline
Bandpass (keV) & 0.1--2  &&0.062--0.21 \\
Energy Res (FWHM)&\multicolumn{2}{c}{45\% at 1 keV}&\\
Area (cm$^2$)&420&80&17.1\\
FOV &2$^\circ$&$ 36 ^{\prime} \times 36 ^{\prime} $&5$^\circ$ \\
FWHM&$ 0.8 ^{\circ} \times 1.7 ^{\circ} $&$ 1 ^{\circ} \times 45 ^{\circ} $\\
 \hline
\end{tabular}
\end{table}

ROSAT was a very successful mission. Main highlights include:
\begin{itemize}
    \item  X-ray all-sky survey catalog, more than 150 000 objects
\item XUV all-sky survey catalog (479 objects)
\item Source catalogs from the pointed phase (PSPC and HRI) containing $\sim$100 000 serendipitous sources
\item Detailed morphology of supernova remnants and clusters of galaxies.
\item Detection of shadowing of diffuse X-ray emission by molecular clouds.
\item Detection (Finally!) of pulsations from Geminga.
\item Detection of isolated neutron stars.
\item Discovery of X-ray emission from comets.
\item Observation of X-ray emission from the collision of Comet Shoemaker-Levy with Jupiter.
\end{itemize}

\subsection{\textsc{Bepposax}}
The SAX Mission (1996 - 2002) was a program of the Italian Space Agency (ASI) with participation of the NIVR. The mission was supported by a consortium of institutes in Italy together with institutes in the Netherlands, and the Space Science Department of ESA.
The acronym SAX stands for 'Satellite per Astronomia X', Italian for 'X-Ray Astronomy Satellite'. SAX was launched on April 30, 1996 by an Atlas Centaur rocket directly into a 600\,km, 96 min orbit at 3.9$^\circ$ inclination. The satellite thus nearly avoided the South Atlantic Anomaly and took full advantage of the screening effect of the Earth's magnetic field in reducing the cosmic ray induced background, an aspect particularly relevant for the high energy instruments. After the successful launch was renamed BeppoSAX in honor of Giuseppe "Beppo" Occhialini. The BeppoSAX mission ended on April 30 2002, and the satellite re-enterd into the atmosphere on April 29, 2003.

The payload of BeppoSAX consisted of five narrow field instruments (NFIs) and a wide field camera (WFC) \cite{Boella_et_al._1997a}. The NFIs consisted of four instruments: 1) the Low Energy Concentrator Spectrometer (LECS), a low energy telescope with a thin window position sensitive gas scintillation proportional counter in its focal plane \cite{Parmar_et_al._1997}; 2) the Medium Energy Concentrator Spectrometers (MECS), a medium energy set of three identical grazing incidence telescopes with double cone geometry, with position sensitive gas scintillation proportional counters in their focal planes \cite{Boella_et_al._1997b}; 3) a collimated High Pressure Gas Scintillation Proportional Counter (HPGSPC) \cite{Manzo_et_al._1997}; and 4) a collimated Phoswich Detector System (PDS) \cite{Frontera_et_al._1997}. The WFC observation were performed by two coded mask proportional counters (PSMPC) \cite{Jager_et_al._1997} that provided access to large regions of the sky. Table \ref{tab:BeppoSAX} reports the main parameters of this instrumentation.
The main feature of BeppoSAX was its broad band, extending from the fraction of keV to more than 100\,keV. This allowed the study of broad band spectra of many classes of galactic and extragalactic objects. 
BeppoSAX produced outstanding results on many classes of galactic and extra galactic sources, however the most spectacular breakthrough of the mission was the discovery of the X-ray afterglows of GRB which allowed the discovery of their optical counterparts confirming the extra-galactic nature of the GRBs \cite{Santangelo_Madonia_2014,paradijs_et_al._1997}.

\begin{table}[h]%
\caption{BeppoSAX.  \label{tab:BeppoSAX}}
\centering

\begin{tabular}{l c c c c c} 
\hline

  &\multicolumn{4}{c}{NFI}&WFC\\
 Instrument & LECS& MECS& HPGSPC&PDS&PSMPC\\

\hline
Bandpass (keV) & 0.1--10&1.3--10&3--120&15--300&2--30 \\
\multirow{2}{*}{Energy Res (FWHM)}& 32\% at 0.28 keV&\multirow{2}{*}{8\% at 6 keV}&\multirow{2}{*}{ 10\% at 6 keV}&\multirow{2}{*}{15\% at 60 keV}&\multirow{2}{*}{20\% at 60 kev}\\
&8.8\% at 6 keV&&&&\\
\multirow{2}{*}{Area (cm$^2$)}&22 @0.25 keV&150 @6 keV&\multirow{2}{*}{450}& \multirow{2}{*}{640}&\multirow{2}{*}{600}\\
&50 @6 keV&101 @8 keV\\
Angular Resolution &3.5$^\prime$ @0.25 keV& 2.1$^\prime$ @6 keV&\multicolumn{2}{c}{collimated}&5$^\prime$\\
FOV (FWHM) 
&&0.5$^\circ$&$1^{\circ}\times1{^\circ}$&1.3$^\circ$&$27^{\circ}\times27{^\circ}$\\
 \hline

\end{tabular}

\end{table}
\subsection{\textit{The program in Japan}}
 \subsection{\textsc{asca}}
    The satellite ASCA, Japanese for "flying bird", \textit{Asuka}, and also an acronym for Advanced Satellite for Cosmology and Astrophysics (ASTRO-D before launch) was launched on February 20, 1993 and operated until July 15, 2000. Its orbit was a near circular one with perigee and apogee distances of 523\,km and 615\,km respectively and an inclination of 31.1$^\circ$. It was the first satellite that used Charged Coupled Devices (CCDs) detectors for X-ray astronomy. It combined imaging capability with broad band-pass, good spectral resolution and large effective area \cite{Inoue_1993}. The scientific payload consisted of 4 identical grazing-incidence X-ray telescope (XRT) each composed of 120 nested gold-coated aluminum foil surfaces and equipped with an imaging spectrometer at its focal plane. Two of the focal detectors were CCD cameras (Solid Imaging Spectrometers or SIS), the other two scintillation proportional counters (Gas Imaging Spectrometer or GIS) \cite{Tanaka_Inoue_Holt_1994}. Table \ref{tab:Asca} summarizes the characteristics of the instruments on board of ASCA.
    
    \begin{table}[h]%
\caption{ASCA (ASTRO-D). The satellite carried 4 x-ray telescopes (XRT). \label{tab:Asca}}
\centering
\begin{tabular}{l c c c c} 
\hline

  \multirow{2}{*}{Instrument} &  \multicolumn{2}{c}{XRT} &SIS&GIS\\
&at 1 keV&at 7 keV&&\\
\hline
\vspace{3.0mm}
Bandpass (keV) & \multicolumn{2}{c}{0.5-12}  &0.4-12&0.8-12  \\
\vspace{3.0mm}
Energy resolution &&&2\% at 5.9 keV& 8\% at 5.9 keV\\
\vspace{3.0mm}
Area (cm$^2$)&1200&600&105&50\\
\vspace{3.0mm}
Angular resolution&\multicolumn{2}{c}{3$^{\prime\prime}$ FWHM} &30$^{\prime\prime}$& 0.5$^\prime$ at 5.9 keV\\
\vspace{3.0mm}
FWHM&$ 24 ^{\prime}$&$ 16 ^{\prime}  $&&\\
 \vspace{3.0mm}
 FOV&\multicolumn{2}{c}{24$^{\prime}$ at 1 keV}&22$^\prime\times$22$^\prime$& 50$^\prime$\\
 \hline
\end{tabular}
\end{table}
    \subsection{\textsc{suzaku}} 
    ASTRO-E2 \cite{Mitsuda_et_al._2007} was the successor of ASTRO-E lost after launch on February 10, 2000. After its successful launch on July 10, 2005, ASTRO-E2 was renamed Suzaku after the mythical Vermilion bird of Asian mythology, the guardian of the southern skies. The mission ended on September 2, 2015. It was developed at ISAS in collaboration NASA/GSFC. It entered a circular orbit at 570\,km altitude with an inclination angle of 31$^\circ$. The orbital period was about 96 minutes. Suzaku carried five Wolter type-I X-ray telescopes (XRT) \cite{Serlemitsos_et_al._2007}. One of them, XRT-S with a focal length of 4.5, featured as X-Ray Spectrometer (XRS) on the focal plane the first X-ray microcalorimeter flown on an orbiting observatory. The X-ray microcalorimeters, the Adiabatic Demagnetization Refrigerator (ADR), and the liquid helium tank were supplied by the NASA/GSFC. The solid neon tank surrounding the helium tank and the mechanical cooler to cool the neon tank were built by ISAS/JAXA. XRS was expected to have an unprecedented energy resolution of 6 - 7\,eV (FWHM) over the 0.3 - 12\,keV energy band. Unfortunately the XRS failed about 29 days after launch, due to the malfunction of the cooling system and the consequent complete loss of liquid He. The other four (XRT-Is), with a focal length of 4.75 m, were devoted to the X-ray Imaging Spectrometer (XIS). The scientific payload of Suzaku was completed by the non-imaging, collimated Hard X-ray Detector (HXD). These instruments operated for ten years, until the end of the mission \cite{Kelley_et_al._2007}. The X-ray Imaging Spectrometer (XIS) was composed of four X-ray sensitive imaging CCD cameras (three front-illuminated and one back-illuminated) of unprecedented energy resolution, each accommodated at the focus of the dedicate telescope (\textsc{xrt-i0}, \textsc{xrt-i1}, \textsc{xrt-i2}, \textsc{xrt-i3}) \cite{Koyama_et_al._2007}. The HXD was a combination of silicon \textsc{pin} diodes and \textsc{gso} scintillators \cite{Takahashi_et_al._2007}. All main characteristics of the SUZAKU instruments are summarized in Table \ref{tab:Suzaku}.
   \begin{table}[h]%
\caption{SUZAKU (ASTRO-E2) carried onboard the XRS (failed after a few weeks), the XIS, and the HXD.}  
\label{tab:Suzaku}
\centering
\begin{tabular}{l c c c c c } 
\hline

  \multirow{4}{*}{Instrument} &  \multicolumn{3}{c}{XRT} &&\\
&XRT-S &\multicolumn{2}{c}{XRT-I}&&\\
&XRS&\multicolumn{2}{c}{XIS}&\multicolumn{2}{c}{HXD}\\
&&\textsc{fi}&\textsc{bi}&\textsc{pin}&\textsc{gso}\\
\hline
\vspace{4.0mm}
Angular resolution& \multicolumn{3}{c}{$\sim 2^\prime$}&&\\
\vspace{3.0mm}
Bandpass (keV) & 0.3-12  &\multicolumn{2}{c}{0.2-12}&\multicolumn{2}{c}{10-60}  \\
\vspace{3.0mm}
Area (cm$^2$)&190&340&390&$\sim$145&315\\
\vspace{3.0mm}
Energy resolution (eV)&$\sim$6.5 at 6 keV&\multicolumn{2}{c}{130 at 6 eV}&3$\times$10$^3$ FWHM&7.6/(E MeV) 0.5 \% FWHM\\
\vspace{3.0mm}
FOV&2.9$^{\prime}\times$2.9$^{\prime}$&\multicolumn{2}{c}{18$^{\prime}\times$18$^{\prime}$}&34$^{\prime}\times$34$^{\prime}$&4.5$^{\circ}\times$4.5$^{\circ}$\\
 
 \hline
\end{tabular}
\end{table} 
    
    \subsection{\textsc{hitomi}} 
    HITOMI, previously ASTRO-H and NEw X-ray Telescope (NEXT), was the result of a international collaboration led by JAXA, the Japanese Aerospace Exploration Agency, that involved many institutions, among others the NASA/GSFC, ESA, CSA, SRON. Hitomi means eye's pupil and probably symbolizes a black hole. The main scientific goal of Hitomi was, namely, the exploration of the structure of the universe: galaxy clusters, black holes, the formation of heavy elements, and the study of physics in extreme conditions with the use of high-resolution spectroscopy combined with a wide-band energy coverage \cite{Takahashi_for_HitomiTeam_2018}.
    Hitomi was equipped with a Soft X-ray spectrometer (SXS), a Soft X-ray imager (SXI), a hard X-ray imager (HXI), and two non-focusing soft Gamma-ray detectors (SGDs). The SXS consisted of a microcalorimeter array of 36-pixel system with an energy resolution $<$7 eV between 0.3 and 12 keV \cite{Kelley_et_al._2016}. The SXI used 4 CCDs cameras \cite{Tanaka_et_al._2018}. The HXI consisted of four layers of 0.5\,mm thick double-sided silicon strip detectors and one layer of newly developed 0.75\,mm thick CdTe double-sided cross-strip detector \cite{Nakazawa_et_al._2018}. The SGD was developed to measure soft $\gamma$-rays via reconstruction of Compton scattering in the semiconductor Compton camera \cite{Tajima_et_al._2018}. These instruments together covered a wide energy range between 0.3--600 keV. The science payload of the mission is summarized in Table \ref{tab:Hitomi}. 
    
    Hitomi featured four optics: two units for the SXT with focal length 5.6\,m, one pointed for the SXS, and the other for the SXI; the other two units were hard X-ray telescopes of 12\,m focal length and had at their focal plane the HXIs.
    
    \begin{table}[h]%
\caption{HITOMI (ASTRO-H) consisted of 2 SXTs and 2 HXTs.}  
\label{tab:Hitomi}
\centering
\begin{tabular}{l c c c c } 
\hline

 Instrument &  SXS &SXI&HXI&SGD\\

\hline
\vspace{2.0mm}

Bandpass (keV) & 0.3-12  &0.4-12&5-80&60-600 \\
\vspace{2.0mm}
Energy Res. (FWHM) (keV at keV)& $<$7$\times$10$^{-3}$ at 6&$<$0.2 at 6&$<$2 at 60&$<$4 at 60\\
\vspace{2.0mm}
\multirow{2}{*}{Area (cm$^2$ at keV)}&300 at 6 &350 at 6 &300 at 30 &$>$ 20 at 100 \\
&250 at 1&370 at 1 &&\vspace{2.0mm}\\

Effective FOV &3$^{\prime}\times$3$^{\prime}$&38$^{\prime}\times$38$^{\prime}$&9$^{\prime}\times$9$^{\prime}$&0.6$^{\circ}\times$0.6$^{\circ}$\vspace{2.0mm}\\
Time resolution ($\umu$s)&5&4/2/0.5/0.1 sNOTA&25.6&25.6\vspace{2.0mm}\\
Angular resolution &$\sim$1.2$^{\prime}$&$\sim$1.3$^{\prime}$&1.7$^{\prime}$ at 30 keV&\\
 \hline
\end{tabular}
\end{table} 
The mission was launched on February 17, 2016 and contact was lost on March 26, 2016. This was due to a series of incidents with the attitude control system leading to an uncontrolled spin rate and breakup of weak mechanical elements.

\subsection{The program in Russia and India}

\subsection{\textsc{granat}}
 
The Granat mission was launched on December 1, 1989 into a highly eccentric
98-hour (4-day) orbit which enabled long uninterrupted observations. The perigee and apogee of the orbit were 2 000 and 20 0000\,km, respectively, and the inclination was 51.9$^\circ$. After an initial period of pointed observations, Granat went into survey mode in September 1994 and operated until November 27, 1998.  It carried seven X-ray and gamma-ray instruments \cite{Roques_et_al_1990}. 

ART-S \cite{Sunayev_et_al_1990}, a large area (4$\times$ 625~cm$^2$) system  of four high-pressure xenon multiwire proportional counters (MWPCs) reached $\sim$ 100 keV with an energy resolution of $\sim$10$\%$ at 60~keV. The collimators of the telescope could work in rocking mode, ON and OFF the target, which enabled the study of timing and spectral parameters of X-ray sources in the pointing mode.
A similar system, ART-P \cite{Sunayev_et_al_1990}, with four identical modules contained position sensitive MWPC together with a coded mask, which enabled imaging with 6$^{\prime}$ resolution in the energy range 4-60\,keV. In the range 4-100\,keV the telescope worked as a spectrometer. The geometric sensitive area of ART-P was 4$\times$630~cm$^2$, the field of view $1.8^\circ\times1.8^\circ$ FWHM. 

The other principal instrument, SIGMA \cite{Paul_et_al_1991}, provided by France, was primarily a gamma-ray instrument (30-1300 keV), but with hard X-ray sensitivity. It featured a coded mask of 13$^{\prime}$ resolution and a position-sensitive scintillation system (PSD, \cite{Roques_et_al_1990}).
The coded aperture, located 2.5\,m from the PSD was an
array of 49$\times$53 square elements, whose basic pattern was a
29$\times$31 Uniformly Redundant Array (URA) \cite{Fenimore_1980}. The opaque 1.5 cm thick tungsten mask elements were bonded to a honeycomb plate that
supported and stiffened the mask assembly without hindering
the transparency of the open mask elements. The dimensions
of the URA mask cell (9.4 mm $\times$ 9.4 mm) imposed the following
key properties of the telescope:
\begin{itemize}
\item the maximum sensitivity rectangular field of view:
$4.3^\circ\times4.7^\circ$, surrounded by a wider field of decreasing sensitivity such that the half-maximum sensitivity boundary was a $10.9^\circ\times11.5^\circ$ rectangle
\item the total detection area: the 794~cm$^2$ central
rectangular zone of the PSD whose size matched the
basic 29$\times$31 mask pattern
\item the intrinsic angular resolution: 13$^{\prime}$
\item the point—source location accuracy: less than
2$^{\prime}$ taking into account the PSD coding element
size (1.175 mm$\times$1.175 mm).
\end{itemize}

An all-sky monitor, WATCH, designed at the Danish Space Research
Institute \citep{Lund_1985,Brandt_et_al_1990}, consisted of four
rotating modulation collimator systems with phoswich detectors for the
purpose of monitoring persistent X-ray sources and localizing transient
events in a significant part of the sky \cite{Brandt_et_al_1990}. 

The detector consisted of parallel NaI (Tl) and CsI(Tl) scintillator strips.  The energy range was from 6 to 180\,keV for NaI and from 20 to 180\,keV for CsI. 
The energy resolution of the detector was modest, about 30$\%$ FWHM at 60\,keV. 

A substantial gamma-burst capability was also on board, including sensitivity in the X-ray and optical regions. 

The PHEBUS instrument \cite{Barat_et_al_1988}, developed by CESR (Toulouse) to record high energy transient events in the range 100\,keV to 100\,MeV, consisted of two independent detectors. Each detector consisted of a bismuth germanate (BGO) crystal 78 mm in diameter by 120 mm thick, surrounded by a plastic anti-coincidence jacket. The two detectors were arranged on the spacecraft so as to observe 4$\pi$ steradians. The burst mode was triggered when the count rate in the 0.1-1.5 MeV energy range exceeded the background level by 8 sigma in either 0.25 or 1.0 seconds. There were 116 energy channels.

The KONUS-B instrument\cite{Mazets_and_Golenetskii_1987}, designed by the Ioffe Physico-Technical Institute in St. Petersburg, consisted of seven detectors distributed around the spacecraft to detect photons from 10\,keV to 8\,MeV. The detectors consisted of NaI(Tl) scintillator crystals 200 mm in diameter by 50 mm thick behind a Be entrance window. Spectra were taken in two 31-channel pulse height analyzers (PHAs), of which the first eight were measured with 1/16 s time resolution and the remaining with variable time resolutions depending on the count rate. The range of resolutions covered 0.25 to 8\,s. The KONUS-B instrument operated from December 11, 1989 until February 20, 1990. Over that period, the 'on' time for the experiment was 27 days. Some 60 solar flares and 19 cosmic gamma-ray bursts were detected.

The French TOURNESOL instrument consisted of four proportional counters and two optical detectors. The proportional counters detected photons between 2\,keV and 20\,MeV in a 6$^{\circ}$$\times$6$^{\circ}$ field of view. The visible detectors had a field of view of 5°$\times$5°. The instrument was designed to look for optical counterparts of high-energy burst sources, as well as performing spectral analysis of the high-energy events. Table \ref{tab:Granat} reports the principal characteristics of the experiments onboard Granat.

Over the initial four years of directed observations, Granat observed many galactic and extra-galactic X-ray sources with emphasis on the deep imaging and spectroscopy of the galactic center, broad-band observations of black hole candidates, and X-ray novae. After 1994, the observatory was switched to survey mode and carried out a sensitive all-sky survey in the 40 to 200 keV energy band.

Some of the highlights included: A very deep imaging of the galactic center region \cite{Mandrou_1990}; The discovery of electron-positron annihilation lines from the galactic microquasar 1E1740-294  \cite{Bouchet_et_al_1991} and the X-ray Nova Muscae \cite{Gilfanov_et_al_1991}; Study of spectra and time variability of black hole candidates.

\begin{table}[h]%
\caption{Granat.The payload consisted of seven experiments: 
\label{tab:Granat}}
\centering
\begin{tabular}{l c c c c | c c c} 

\hline
&&&&&\multicolumn{3}{c}{$\gamma-$ray burst esperiment }\\
Payload & ART-S  & ART-P  &SIGMA & WATCH&PHEBUS& KONUS-B & TOURNESOL \\

\hline
Bandpass (keV) &3-100 &4-60$^a$  &30-1300 &6-180&&& \\
Bandpass (MeV) & &  & & &0.1-100&0.02-8& 0.002-20\\
Area (cm$^2$)&2400 &1250 &800 &45 &100 &315 &\\
Ang.Resol.&&6$^{\prime}$&13$^{\prime}$&&&&\\
Field of View (FOV)&$2^\circ\times2^\circ$&$1.8^\circ\times1.8^\circ$& $5^\circ\times5^\circ$&All sky&All sky& All sky&$5^\circ\times5^\circ$ \\
 Ener.Res. at 5.9 keV&21$\%$&25$\%$&&&&&\\
 Ener.Res. at 22.1 keV&13$\%$&&&&&&\\
 Ener.Res. at 59.5 keV&11$\%$&14$\%$&&30$\%$&&&\\
 Time Resol. ($\mu$s) &200&&&&&&\\
 
 \hline
\multicolumn{8}{l}{$^a$ Notes: for imaging. The energy range for spectroscopy and timing was 4-100\,keV}\\

\end{tabular}
\end{table}

\subsection{\textsc{irs-p3}}
IRS-P3 was the sixth satellite in Indian Remote Sensing satellite series, an Earth observation mission launched under the National Natural Resources Management System programme (NNRMS) undertaken by ISRO. The launch of IRS-P3 took place on March 21, 1996, its orbit was initially an elliptical one, but after few weeks it was stabilized in a sun-synchronous circular orbit with an altitude of 817 km, an inclination of 98.7$^\circ$, a repeat cycle  of 24 days, and a period of 101.35 min. The objectives of the mission was the processing and interpretation of data generated by its two instruments, the Wide-Field Sensor (WiFS) and Modular Opto-electric Sensor (MOS), developed by the German Aerospace Center (DLR). It also hosted a scientific instrument, the Indian X-ray Astronomy Experiment (IXAE), for the study of X-ray astronomy. The payload was designed to study periodic and aperiodic intensity and spectral variations in galactic and extra-galactic X-ray sources like pulsars, X-ray binaries, Seyfert galaxies, quasars etc., and to study light curves and spectral evolution of transient phenomena. The instrumentation consisted of a Pointing Proportional Counter (PPC) and an X-ray Sky Monitor (XSM) \cite{Thyagarajan_Neumann_and_Zimmermann_1996}. The principal characteristic of the X-ray astronomy experiment are summarized in Table \ref{tab:IRS-P3}

\begin{table}[h]%
\caption{IRS-P3 carried on board two instrumentation to conduct X-ray astronomy: a Pointing Proportiona Counter (PPC) and an X-ray Sky Monitor (XSM)  \label{tab:IRS-P3}}
\centering
\begin{tabular}{l c c }
\hline
Experiment&PPC&XSM\\
\hline
Energy Range (keV) &2-20 &2-8 \\
Area (cm$^2$)&1.1$\times$1.1 &1\\
Field of View (FOV)&$2^\circ\times2^\circ$&$90^\circ\times90^\circ$\\

 \hline

\end{tabular}
\end{table}

\section{Conclusions}

With this chapter, we wanted to summarise the history of X-ray Astronomy from its inception until around the end of the 1990s. We have chosen for this contribution an approach that allows the reader to find the essential data, and references for X-ray astronomy missions (be they rockets, balloons, or satellites). The history of X-ray astronomy has obviously not stopped and the new millennium has seen the launch of new missions that are still active, Chandra, XMM-Newton, and ASTROSAT to name but a few. There are specific chapters on these and other missions in the X-ray section of the Handbook. 
We would like to conclude this chapter with a tribute to the most significant figure in the history of Astronomy X: Riccardo Giacconi. There is no doubt that Bruno Rossi is to be credited with the initial intuition of attempting X-ray observation of the sky. However, it was Riccardo Giacconi who gave substance to that idea from the very first visionary studies of X-ray telescopes, effectively opening up a new Astronomy. It is therefore no coincidence that he was awarded the Nobel Prize in 2002 with the motivation "for pioneering contributions to astrophysics, which have led to the discovery of cosmic X-ray sources". Not only did Giacconi open up the vision of the X-ray sky, but as Josh Grindlay noted\footnote{In his seminar at the Memorial Symposium to Honor Riccardo Giacconi, held at the National Academy of Sciences in Washington, DC on May 29-30, 2019.} "In fact he not only opened the entire field of X-ray Astronomy/Astrophysics but also opened the then new field of Time domain Astrophysics. The broad  phenomenology of Bursts (X-ray, Gamma-Ray, now Radio…) was not part of the Astronomical Landscape until dramatically variable sources like Cyg X-1 and “X-ray Novae” were discovered.  Then, GRBs, enormous Flaring from Blazars,  BH-LMXB outbursts demanded physical understanding! Exciting new Astrophysics, continuing through today". 
In the coming years, new missions will be launched such as SVOM, EP, and XRISM. Other missions such as eXTP and ATHENA are at an advanced stage of study. We are therefore confident that this chapter will grow over time, enriched with new stories. 

\section{Cross-References}
Finn E. Christensen \& Brian D. Ramsey 'X-ray Optics for Astrophysics: a historical review'

Thomas Siegert, Deirdre Horan, Gottfried Kanbach 'Telescope Concepts in Gamma-Ray Astronomy'

Enrico Costa 'General History of X-Ray Polarimetry in Astrophysics'

\section{Acknowledgements}
The authors sincerely thank Dr. Erik Kuulkers of ESA for the thorough review of the manuscript and for providing many interesting suggestions and comments.

\newpage
\section{Appendix 1. List of the rockets launched from 1957 to 1970. }
\begin{enumerate}
\item
\textbf{February 13, 1957} First Skylark test flight from Woomera. The rocket didn't carry any scientific payload
\item 
\textbf{November 13, 1957} 
(Skylark program, first flight at Woomera in February 
payloads of 150 kg to 300 km (10 mins for x-ray observation)
    \item
\textbf{July 08, 1959} British National Program. Skylark SL14 from Woomera Australia, carrying a pair of X-ray cameras, was launched at Woomera on 1959 July 8, attaining a peak altitude of 93 km. The film used was a Kodirex emulsion prepared by Dr. Herz at Kodak Ltd. The integrated solar fluxes in the range 2-8 {\AA} and 8-20 {\AA} \cite{Pounds_2010}
Unfortunately, parachute recovery failed and both cameras were destroyed on impact \cite{Pounds_and_Bowen_1962}.
\item
\textbf{September 17, 1959} British National Program. Skylark SL12 from Woomera Australia, attaining a
peak altitude of 132 km. Again the parachute failed to operate properly and the vehicle impacted at some 1.3 km/sec.\cite{Pounds_and_Bowen_1962}
\item
\textbf{June 18, 1962} the first rocket of the AS\&E-MIT group was launched from White Sand (New Mexico). The payload consisted of three large area Geiger counters aimed at the detection of solar X-ray reflected from the moon surface. As well known and as discussed above, this kind of radiation was not found but, having excluded all other possibility, the authors concluded: "However, we
believe that the data can best be explained by
identifying the bulk of the radiation as soft x rays
from sources outside the solar system \cite{Giacconi_et_al._1962}." For correctness is fair to say that the authors never wrote in the discovery paper that the source was located in the Scorpio constellation, they only cited the peculiar objects Cassiopeia A and Cygnus A located within the field of view of the instruments.
\item
\textbf{September 30, 1962}, Fisher and Meyerott launched the first of two Lockheed Aerobee rockets 4.69 and 4.70 (s. below) from Wallops Island, Virginia. The authors suggested the existence of many region of the sky in which emissions in the energy band of X-ray exist. The rockets had onboard several gas proportional counters with energy range from 3 to 20 keV \cite{Fisher_and_Meyerott_1964a}.
 
\item
\textbf{October 12, 1962}. The rocket by AS\&E-MIT did not observe any peak in the flux. However, a different sky region than in June 62 had been observed. The observational data were collected by three Geiger counters with effective area of 10 cm$^2$ arranged in anti-coincidence mode \cite{Gursky_et_al._1963}.
\item
\textbf{March 15, 1963}. The Aerobee rocket 4.70 was launched to locate stellar sources of soft X-rays and obtain information about their spectra. The  payload consisted of several well collimated ($5^{\circ}\times10^{\circ}$) gas proportional counters that operated in the energy range 0.2--20 keV \cite{Fisher_and_Meyerott_1964a}. A careful re-analysis of the data collected during this and the previous Aerobee launch of Fisher's group, confirmed the existence of the X-ray emitting regions but with a very low statistical evidence \cite{Fisher_et_al._1964}. Also the failure in detecting the source in the region of the Crab Nebula was revised \cite{Bowyer_1964,Fisher_and_Meyerott_1964b}.
\item
 \textbf{April 29, 1963}. NRL launched an Aerobee rocket, carrying a proportional counter of 65 cm$^2$ effective area, from White Sands, New Mexico. The counter was sensitive in the 1--8 {\AA} range with a FWHM of 10$^\circ$. The observations confirmed with better localization the source in the constellation Scorpius and gave also evidence of a new source in the direction of the Crab Nebula. Hypothesis of the nature of the sources, i.e. neutron stars remnant of supernovae explosions, were advanced
 \cite{Bowyer_et_al._1964a, Bowyer_et_al._1964b}. 
 \item
\textbf{June 11, 1963} A new rocket by AS\&E-MIT observed of at least three different extra solar X-rays sources: Sco X-1, Cyg X-1 and Crab, which were identified in the published paper as N1, N2 and N3 \cite{Gursky_et_al._1963}. The payload consisted of three Geiger counters of 10 cm$^2$ effective area each and Argon as filling gas. Anticoincidence was used to reduce the cosmic ray background.
\item
\textbf{June 16, 1964}. NRL surveyed the galactic plane from the southern part of Scorpius through Cygnus to the norther part of Perseus. It carried onboard two Geiger counters sensitive in the wavelength range 1--15 {\AA}, and with large effective area of 906 cm$^2$ \cite{Bowyer_et_al._1965}. The detected sources were located near the galactic plane. The detectors were two Geiger counters sensitive in the wavelength range 1--15 {\AA}  and of a large effective area of 906 cm$^2$ \cite{Bowyer_et_al._1965}.
\item
\textbf{July 7, 1964}. NRL confirmed the localization of the source in the Crab Nebula (called Tau X-1!), with an experiment performed during the lunar occultation of the Crab. The Aerobee rocket carried two Geiger counters with a Mylar transmitting window of 114 cm$^2$ \cite{Bowyer_et_al._1964c}
\item
\textbf{August 28, 1964}. AS\&E-MIT confirmed the detection of Sco X-1 and observed a new extended source in the Sagittarius, coincident with SgrA, the radio centre of the Galaxy. The detector device was a bank of Ar-filled Geiger tubes, its total effective area was 70 cm$^2$, and the sensitivity range 1--9 {\AA}. A novel collimator with high angular resolution and wide field of view (FWHM 15$^{\circ}\times 21.5^{\circ}$) was accommodated in front of the detector \cite{Giacconi_et_al._1964}, \cite{Oda_et_al._1965}.
\item
\textbf{October 1, 1964}. The launch was performed by Fisher and colleagues of the Lockheed Missile and Space Center. The instrumentation on-board consisted of five gas-filled (90\% argon and 10\% methane) proportional counters and collimators and three photometers for detecting visible starlight. Three of the five proportional counters had large-area windows of $\sim263$ cm$^2$, whereas the remaining had a smaller effective area of 18 cm$^2$. The X-ray detectors were sensitive in the energy range between 2 and 20 keV. They observed 8 discrete sources, probably located in the Galaxy. No radio or optical counterpart were identified. Suggestion that the flux of the sources may change with the time \cite{Fisher_et_al._1966}.
\item
\textbf{October 26, 1964}. The AS\&E-MIT team constrained with better angular resolution the position of Sco X-1, Sco X-2 and Sgr X-1. In contrast with NRL group no X-source in SN1604 was observed The on-board instruments was essentially as in the fly of August 28th, 1964 \cite{Clark_et_al._1965},\cite{Oda_et_al._1965}.

\item \textbf{November 25, 1964.} NRL's Aerobee launch to perform a Sky Survey from Perseus to Puppis. The only source detected was the Crab Nebula. The same equipment as in the fly of the 16th of June was used, however during this flight the counters were arranged in an anticoincidence mode to reject the cosmic ray background \cite{Bowyer_et_al._1965}.

\item \textbf{March 17, 1965}. AS\&E-MIT Aerobee launch from White Sand. The rocket carried on-board (for the first time) a grazing incidence imaging telescope, developed by Giacconi \cite{Gacconi_and_Rossi_1960} (Wolter type). An imagine of the X-ray emission from the Sun was obtained \cite{Giacconi_et_al._1965}. 

\item\textbf{April 25, 1965.} NRL's survey report 29 sources. The survey allowed search for variability for sources such as Cyg X-1 and Cyg X-2. The detectors were Geiger counters filled with gas \cite{Friedman_Byran_Chubb_1967}.

\item \textbf{September 30, 1965.} the Lockheed Company supported by NASA launched a rocket from White sands. Five X-ray sources at low galactic latitude and their spectra were observed. The sources may correspond to Sco X-2 and Sgr X-2. In the following Table \ref{tab:Rocket_30_09_65} are summarized the characteristics of the two X-ray detectors on-board the rocket. \cite{Fisher_et_al._1967}
\begin{table}[h]%
\caption{Characteristics of the A and B proportional counters on-board the rocket of the Lockheed Company launched on the 30th of September 1965 \label{tab:Rocket_30_09_65}}
\centering
\begin{tabular}{ l l l |l l| l } 

\hline

\multirow{3}{*}{Detector}& \multicolumn{2}{c|}{Beryllium window}&\multicolumn{2}{c|}{Filling Gas}&\multirow{3}{*}{FWHM}  \\
& Average Thickness & Effective Area & Pressure &Mixture\\
&(mgms/cm$^2$)&(cm$^2$)& (mm Hg)&&\\
\hline
\multirow{2}{*}{A}&\multirow{2}{*}{26.7}&\multirow{2}{*}{195}&\multirow{2}{*}{800}&90\% argon&\multirow{2}{*}{$1.7^{\circ}\times16.7^{\circ}$}\\
&&&&10\%methane\vspace{3mm}\\

\multirow{4}{*}{B}&\multirow{4}{*}{40.3}&\multirow{4}{*}{235}&\multirow{4}{*}{1600} & 69.5\% argon&\multirow{4}{*}{$1.8^{\circ}\times7.4^{\circ}$}\\
&&&&10\%methane\\
&&&&20\%xenon\\
&&&&0.5\%helium\\
\hline
Bandpass (keV) &\multicolumn{5}{c}{2.5--16} \\

\hline

\end{tabular}
\end{table}
\item \textbf{October 28, 1965} Two stage Hydra-Iris rocket n from a floating launcher 720 km offshore southern California. The detectors were a thin-window proportional counter and a Phoswich scintillation detector. The proportional counter had a total area of 91 cm$^2$ and was filled with a mixture of Argon (90\%) and Methane (10\%) and had a FOV of 7$^{\circ}\times30^{\circ}$. The phoswich detector was composed of cesium iodide surrounded by plastic scintillators (0.1 cm thick in the front and 0.64 cm in the back and on the sides), the FWHM was 10$^{\circ}\times35^{\circ}$. observation of Sco X-1, Tau X-1 and cyg X-1 were made and their spectra were measured in the energy range 1--40 keV \cite{Grader_et_al._1966}. 

\item \textbf{March 8, 1966.} Oda's modulation collimator to pinpoint Sco X-1 and individuation of the optical counterpart using the data from the Tokyo  Astronomical Observatory and Mount Palomar Observatory  \cite{Gursky_et_al._1966}, \cite{Sandage_et_al._1966}, in the same flight was studied the size and the position of the X-ray source in the Crab Nebula \cite{Oda_et_al._1967}

\item 
\textbf{ September 20 and 22, 1966} from Johnston Atoll by the Lawrence Radiation Laboratory, University of California, (Livermore).  The group conducted study of the background X-radiation between 4 and 40 keV. The first rocket carried "two beryllium-window, xenon-filled, proportional counters" sensible in the region 4--25 keV, and the second "a sodium iodide scintillation counter surrounded by a plastic fluor cosmic-ray anti-coincidence detector" for observations in the range 8--80 keV. \cite{Seward_et_al._1967}

\item 
\textbf{October 11, 1966} AS\&E-MIT. GX3+1 and a survey of the X-ray sources in the  Cygnus region principally. Cyg X-1, Cyg X-2 (confirmed), Cyg X-3,  Cyg X-4 (new sources), CygA no X-ray source. "The X-ray detectors were argon-filled, beryllium window proportional counters with an effective area of about 800 cm$^2$" and they operated in the energy band 1.5--12 keV \cite{G3_and_W_1967}.
 
\item 
\textbf{April 4 and 20, 1967} Skylark rockets, launched from Woomera (Australia). The group of the Adelaide University and Tasmania University initially reported of a strong source in the Crux constellation, a revision of the data suggests the identification of the source with Cen X-2 \cite{Francey_et_al._1967}. The Skylark detectors onboard consisted  of two independent pairs of proportional counters filled with a xenon-methane mixture. The spectral information from the detectors consisted of the counting rates from two energy bands, 2--5 keV and 5--8 keV. The X-ray window of each counter was 12 cm$^2$, and consisted of 14 mg/cm$^2$ of beryllium.
Collimators were placed in front of each of the counters to define its field of view, the angular resolutions being 10.5$^\circ$ FWHM in the direction of rocket spin and 35$^\circ$ FWHM in the plane containing the spin axis \cite{Harries_et_al._1967}. 
 
\item 
\textbf{April 10, 1967} Cooke Leicester University. Skylark SL118 and SL119 from Woomera South Australia. The payload consisted of ``a large-area proportional counter" with an effective area of 295 cm$^2$, sensible in the energy range 2 to 5 keV. The scientific goal was to perform an high-sensitivity survey of the X-ray sources in the southern hemisphere. Three point sources were detected: the already know Sco X-1 with new calculated intensity of 1.3 $\times$
10$^{-7}$erg cm$^2$ sec$^{-1}$; Tau X-1 a much more weaker source in the Taurus region; and the outstanding observation of Cen X-2. This bright source was the same observed by the  Australian scientist during the flights on the 4th and 20 April 1967 as above reported \cite{Chodil_et_al._1967a}. However, when observed five weeks later, on the 18th of May, by the group of the Lawrence Radiation Laboratory group of the California University, Livermore (s. below)  this source appeared six times fainter. Note that on October 28, 1965 the same region of the sky was observed during another rocket flight of the California University (s. above) and the same source (Cen X-2) was ``definitively absent" \cite{Grader_et_al._1966}. Thus for the first time the existence of variable X-ray sources was suggested \cite{Cooke_et_al._1967}.

\item 
\textbf{May 17, 1967} NRL White Sand. Survey of the Virgo region. Signals from the directions of the Quasar 3C273, and of Virgo A (M87). Suggestion that the "diffuse" background emission could be due to unresolved point sources. The detectors were two proportional counters (90\% Argon and 10\% Methane) with a FWHM of $1^{\circ}\times8^{\circ}$ and a total effective area of 525 cm$^2$, the X-ray were detected in the wavelength 1 to 10 {\AA} \cite{Friedman_and_Byram_1967}.

\item 
\textbf{May 18, 1967} Kauai, Hawaii. A Nike-Tomahawk rocket Lawrence Radiation Laboratory, University of California, Livermore, California.  ABSTRACT  "This paper reports the results of x-ray spectrum and location measurements of several cosmic x-ray sources made on 18 May 1967. Rocket-borne proportional counters were used. The x-ray spectra of Sco XR-1, Tau XR-1, and Lup XR-1 were measured, and the location of Vel XR-1, Lup XR-1, and a new source Cen XR-3 were determined. In addition, the x-ray spectrum and location of a variable source, Cen XR-2, were obtained". The payload consisted of two gas filled (90\% Xenon and 10\% Methane) proportional counters sensitive in the energy range between 2 to 30 keV, the effective area was 87 cm$^2$, and the FWHM $10^{\circ}\times30^{\circ}$ \cite{Chodil_et_al._1967a}, \cite{Chodil_et_al._1967b}.

\item 
\textbf{July 7, 1967} MIT measurements of the X-ray sources in the Sagittarius region with a precision of 20'. Discovery of a probable source (the sixth), evidence of X-ray radiation from M87 "Our detectors consisted of two banks of argon-filled (3.8 mg/cm$^2$) proportional counters with beryllium windows of 50$\mu$ nominal thickness" \cite{Bradt_et_al._1967},\cite{Bradt_et_al._1968}.

\item 
\textbf{September 7, 1967} NRL Aerobee from White Sand. Five target Sco X-1, the galactic pole, Cyg X-1, Cyg X-2, the Earth. The rocket carried three proportional counters with FWHM of 10$^\circ$, two had an effective area of 100 cm$^2$ and were sensitive in the soft X-ray range (0.20--0.28 keV), the other one, of 200 cm$^2$  effective area was sensitive in the range 1.5--13 keV \cite{Henry_et_al._1968}, \cite{Fritz_et_al._1968}

\item 
\textbf{September 7, 1967} X-ray air-glow Nike-Tomahawk two stage Rocket from Tonopah (Nevada) California University. Soft X-ray spectra from Crab and Sco X-1. The instrumentation onboard consisted of two thin-window proportional counters filled with a gas mixture of 90\% neon and 10\% methane, one had an area 25 cm$^2$ and the other of 65 cm$^2$, the energy range of experiment was 0.4--5.5 keV. The counters were collimated between $\pm 30^\circ$ vertically and $\pm 5$ horizontally \cite{Grader_et_al._1968}


\item 
\textbf{February 2, 1968} AS\&E-MIT Aerobee 150 same payload as the flight of 11 October 1966. X-ray background measurement in the energy range 3-13 KeV \cite{Gorenstein_et_al._1969} and the location of an X-ray source in the Vela constellation (Vela X-1 e/o GX263+3) \cite{Gursky_et_al._1968}

\item 
\textbf{June 12, 1968} Leicester University. Skylarck rocket (SL723). From Woomera Rocket Range (South Australia). Peak of intensity in the region of the Virgo (Vir X-1). The rocket carried two proportional counters of an effective area of 1,385 cm$^2$ and a FWHM of $28^{\circ}\times8^{\circ}$. One of the counters was sensitive in the energy range 1.4--2.5 keV, and the other in the range 2--18 keV. \cite{Adams_et_al._1969}

\item 
\textbf{July 8, 1968} Leicester University. Skylarck 403 rocket. Originally intended to observe Cen X-2. The rocket took onboard two proportional counters of small effective area (25 cm$^2$) and different FWHM, namely 20$^{\prime}$ and 3.5$^{\circ}$ \cite{Adams_et_al._1969}.

\item 
\textbf{July 26, 1968} from White Sand. MIT group. Upper limit of the angular size of  three X-ray sources in the Sagittarius. The detection system consisted of proportional counters of a total effective area of $\sim $75 cm$^2$ and FOV $1.2^{\circ}\times65^{\circ}$. The energy range was limited between 1.5--6 keV. In front of the counters was placed a modulation collimator consisting of two parallel, plane grids composed of many parallel wires \cite{Polucci_et_al._1970}. 

\item 
\textbf{July 27, 1968} Aerobee 150 from White Sand. Bob Novick Columbia University Search of X-Ray polarization in Sco X-1 NO POLARIZATION Termal Bremsstrahlung. The detector was a Thomson-scattering x-ray polarimeter sensitive in the range 6--18 keV. The scattered radiation was collected in Xe-Methane proportional counters    \cite{Angel_et_al._1969}.

\item 
\textbf{October 29, 1968} from Johnston Atoll by the Lawrence Research Laboratory (LRL) University of California (Livermore). X-rays from Large Magellanic Cloud.The payload consisted of proportional counters filled with a gas mixture of 90\% Argon and 10\% Methane and with an aluminized Mylar window of 323 cm$^2$ effective area, the FWHM was  $2.5^{\circ}\times75^{\circ}$ \cite{Hans_Mark_et_al._1969}.

\item 
\textbf{December 5, 1968} AS\&E-MIT same payload as the flights of the 11 October 1966 and 2 Februar 1968, on-board an Aerobee rocket tentative hypothesis about the sources of X-ray. There were four detector for an effective aera of 800 cm$^2$, the FWHM of the detectors was $2^{\circ}\times45^{\circ}$ although they were arranged to have different field of view. The energy range of the observation was 1 to 10 keV. Survey of the Cassiopea region. Cas A and SN 1572 are X-ray sources. Possible origin of cosmic X-rays in SN remnant, but not all the sources found are SN. There is an inverses proportionality between the radius of SN and the intensity of the X-ray radiation. It would be interesting to measure the spectrum of the radiation (still experimental difficulties) \cite{Gorenstein_et_al._1970} 

\item 
\textbf{March 7, 1969}  Aerobee 150 from White Sand. Bob Novick Columbia University Search of X-Ray polarization in the X-ray emission from Taurus X-1 (Crab) No polarization but it cannot be excluded a polarization of the same magnitude as in radio or optical band. The experiment was conducted by means of a polarimeter that used incoherent scattering, in addition a pulse rise-time discrimination was foreseen in order to reject the cosmic ray background. The energy range studied was 5.5 to 22 keV \cite{Wolff_et_al._1970}.

\item 
\textbf{March 13, 1969} NLR Aerobee White Sand. The rocket payload consisted of two proportional counters both filled with a gas mixture of 10\% methane in argon. The counters differed in effective area and material of the windows Mylar and Teflon, and 300 cm$^2$ and 250 cm$^2$ respectively. The Mylar counter was sensitive in the range 1--13 keV while the Teflon ones between 1.2 to 13 keV. About the scientific success of the mission we report the words of Fritz, Friedman and colleagues \cite{Fritz_et_al._1969}:
\begin{quote}
    [...]discovery of an x-ray pulsar in the general direction of the Crab Nebula.[...] The period of the pulsar is identical with that of the radio and optical pulsar NP 0532, identification of the x-ray pulsar with that object appears certain.
\end{quote}

\item
\textbf{April 1, 1969} from Woomera, Australia. The Leicester University group launched a Skylark rocket (SL724) that carried onboard two proportional counters. Each counter had an effective area of 1,380 cm$^2$ and a FWHM of 27.4$^{\circ}\times$4$^\circ$. Due to the characteristics of the rocket motion, the sources were pointed 15 times longer than the previous Skylark flight (SL723 on 12 June 1968) providing an increased sensitivity. Seven X-ray sources in the region of Centaurus, Norma and Lupus constellations were observed and their spectra were studied \cite{Cooke_and_Pounds_1971}.

\item 
\textbf{April 24, 1970} White Sand Columbia University. Kitt Peak National Observatory, Space Science Division. X-ray spectrum from Sco X-1. The detector was a Bragg spectrometer composed of two large-area (17 inch $\times$9 inch) reflecting crystals of synthetic graphite.The FWHM of the first panel was $0.65^{\circ}\pm0.05^{\circ}$, that of the second $0.75^{\circ}\pm0.05^{\circ}$. The X-rays that satisfied the Bragg conditions were reflected on a bank of proportional counters with beryllium window of effective area 1220 cm$^2$. An independent proportional counter experiment (0.6--30 keV) was posed in the front face of the rocket in order to determine the temperature of the observed source Sco X-1. \cite{Kestenbaum_et_al._1971} \cite{Kestenbaum_Angel_and_Novick_1971}.

\item 
\textbf{June 26, 1970} AS\&E Aerobee Rocket from White Sand. A survey of known soft X-ray source with the new grazing telescope technique and comparison with non-focusing detectors. The payload consisted of two components: a focusing mirror system; and a thin window position sensitive proportional counter. The instrumentation was conceived for observation in the wavelength 10{\AA}  to 80{\AA} and had a FOV of $2^{\circ}\times8^{\circ}$. The effective area of the collector telescope was 160 cm$^2$, while the overall detectors area was 100 cm$^2$ \cite{Gorenstein_et_al._1971}.

\item
\textbf{September 27, 1971} Launch of the Skylark SL1002 from Woomera, the rocket was a Sun-pointing Skylark a new type of vehicle, available since 1964, with a payload attitude control system, that allowed more precise experiments. The launch was carried out during a lunar occultation of the sorce GX 3+1 with the aim of identifying its optical counterpart. No optical counterpart was seen.\\ \textbf{October 24, 1971} One month later, aided by predictions from Leslie Morrison at the Royal Greenwich Observatory, the successive lunar eclipses of GX 3+1 was observed during the Skylark flight SL 974\footnote{The rocket used was a spin-stabilized one (SL 974) launched as usually from Woomera by the Leicester and MSSL reaserchers}, obtaining a source position to within 0.2$^{\prime\prime}\times$5$^{\prime\prime}$, a precision unsurpassed for the next decade \cite{Janes_et_al._1972}. Unfortunately, the tiny error box was empty, with no stellar counterpart identified. Eventually 2012, using the High Resolution Camera on board the Chandra X-ray Observatory, Maureen van den Berg, Jeroen Homan, Joel K. Fridriksson, and Manuela Linares discovered the faint infrared counterpart of GX 3+1 \cite{van_den_Berg_et_al._2014}. Those Skylark flights did, however, show the potential of the lunar occultation technique to aid source identification – a method that later formed the basis of HELOS, Europe’s first X-ray mission.

\end{enumerate}

\newpage
\section{Appendix 2. List of the Balloon missions launched by the MIT group.}

\begin{enumerate}

\item \textbf{July 21, 1964}. Launch from the balloon base of Palestine in Texas. The first  X-ray spectrum of the Crab nebula in the energy range  15--60~keV was observed. The payload consisted of a scintillation counter employing a NaI(TI) crystal with an area of 97 cm$^2$, and 1 mm thick. A brass slats collimator limited the field of view to $\pm16^\circ$ in one direction, and to $\pm55^\circ$ in the other\cite{Clark_1965} .

\item \textbf{July 18, 1966}. Launch from Palestine to perform a survey of the northern sky using a balloon-borne detector with sensitive area of 368 cm$^2$, 12$^\circ$ FWHM field of view, and a direction determination accuracy of $\pm2^\circ$. The instrument had four scintillation X-ray detectors surrounded on the bottom and sides by a graded radiation shield of lead and tin inside of a plastic scintillator anticoincidence shield. Each detector was a NaI(T1) crystal (1 mm thick, 92 cm$^2$ effective area) optically coupled to a 5-inch photomultiplier by a cylindrical Lucite light pipe. The spectra in the range 20-100~keV of Tau X-1 and Cyg X-1, and upper limit on the intensity change of Cyg X-1/Cyg X-3 were reported \cite{Clark_et_al._1968}.

\item \textbf{September 19, 1966} from Palestine. Scientific goals of the mission were the study of the X-ray emission from Cyg X-1, from Cassiopea A, and from the Tycho's supernova remnant (SN 1572) in the energy range 20--100 keV. The detector was a high resolution NaI-scintillation counter with 56.3 cm$^2$ effective area, and 8.4$^\circ$ FOV \cite{Overbeck_et_al._1967}  

\item \textbf{Februar12, 1967} from Page, Arizona. Survey of the northern sky in the range 20-100 KeV, the detector was an actively shielded 400 cm$^2$ NaI(T1) spectrometer \cite{Lewin_et_al._1967}. 

\item \textbf{May-June 1967}. Three balloons were flown on the 15th and the 24th of May, and on the 26th of June from Palestine (Texas). Variation of ScoX-1 and Cyg X-1 were reported. The payload was the same used during the flight of the 19th of September 1966 \cite{Overbeck_and_Tananbaum_1968a}, \cite{Overbeck_and_Tananbaum_1968b}

\item \textbf{October 1, 1967} from Palestine. The scientific goal was the investigation of X-ray sources in the Crab Nebula. The payload was a set of proportional counters filled with a mixture of 90\% xenon in 10\% nitrogen of a total area of 5000 cm$^2$. The energy band of the observations was 20 to 70 keV \cite{Glass_1969}

\item \textbf{October 15, 1967} from Mildura (Australia). The instrumental payload consisted of four NaI(T1) scintillation detectors with a total sensitive area of 358 cm$^2$ and an anticoincidence jacket of plastic scintillators. The field of view had an angular width of 13$^\circ$ FWHM. The observed X-ray energies ranged between 20 and 105 keV. The mission provided the observation of high energy X-ray sources in the southern sky \cite{Lewin_et_al._1968a} and observation of an X-ray flare from Sco X-1 \cite{Lewin_et_al._1968b}. 

\item \textbf{October 24, 1967} from Mildura. Exploration of the Large and Small Magellanic Clouds searching for X-ray sources. In this flight the detector consisted of four NaI(T1) scintillation detectors with a total sensitive area of 368 cm$^2$, an energy range 20--110 keV and a FWHM of 17.5$^\circ$. An anticoincidence jacket of plastic scintillator completed the experimental assembly. No evidence was found of discrete sources in the directions of the large and small Magellanic clouds \cite{Lewin_et_al._1968c}.

\item \textbf{April 24, 1968} from Palestine. X-ray flux from Scorpio X-1 
Flight duration was about 13 hours. 

\item \textbf{July 16, 1968} from Palestine. Observation of Cyg X-1 in the energy range 25 to 100 keV. The instrumentation comprised two X-ray telescope with gyro-stabilized modulation collimators. The collimated X-ray were detected by two NaI(T1) scintillators of 280 cm$^2$ effective area and a FWHM of $10^{\circ}\times10^{\circ}$ \cite{Floyd_1969}.

\item \textbf{October 25, 1968} from Palestine (Texas). Study of high energy X-ray from M87. The dector consisted of two xenon proporional counters with a total effective area of 2000 cm$^2$, they operated in the energy band 15--65 keV with a FWHM of $13^{\circ}$ \cite{McClintock_et_al._1969}.

\item \textbf{March 20, 1969} from Mildura (Australia) MIT. Southern sky survey, decrease of the X-ray intensity in Cen X-2, Periods and continual changes in Sco X-1. The payload was more or less the usual ones, namely the  358 cm$^2$ scintillator surrounded by an anticoincidence arrangement of plastic scintillators, with a FWHM of $13^{\circ}$. The experiment was conducted in the energy range 20 to 100 keV \cite{Lewin_et_al._1970}, \cite{Lewin_McClintock_Smith_1970}.

\item \textbf{April 16, 1969} from Mildura. Lewin et al. A flare in a source (Crux) not emitting during the observation on 20 March 1969, but present in the data of the balloon flight on the 15Th of October 1968. The balloon, that was lost in the Tasmanian see after $\sim$ 16 hours , reappeared nine months later in New Zealand. Thanks to the expertise of the technicians of Kodak company the data registered were made legible although with poor quality. As a consequence, only three of the seven detection channels were read. Therefore the energy range examined was between 18 and 36 keV (originally 18--100 keV). The experimental apparatus was the same of the previous launch on March 20 \cite{Lewin_et_al._1971a}.

\item \textbf{May 10, 1969} Palestine,  Hard X-ray from Crab nebula. The measurement covered energies of 25 to 100 keV. The instrument consisted of two NaI(T1) proportional counters with anticoincidence shield, put behind a modulation collimator. The total effective area was 140 cm$^2$  \cite{Floyd_et_al._1969}

\item \textbf{October 15, and 16, 1970} from Mildura.  Flare in the sources GX 301-2 and GX 304-1, first detection of the X-ray source GX1+4 \cite{Lewin_et_al._1971b}. The measurements were made using a 45 cm$^2$ NaI(T1) scintillation detector, surrounded by a NaI(T1) anticoincidence jacket. The slit field of view had an angular width of $1.5^ {\circ}\times 13^ {\circ}$ FWHM. Eight energy channels recorded X-ray in the range 15 to 150 keV \cite{Lewin_et_al._1972}.

\item \textbf{April 5, and 6, 1972} from Alice Spring (Australia).During the 24 hours flight observation of the Large and Small Magellanic Clouds were performed as well as of the regions along the galactic plane, The payload consisted of NaI(T1) scintillator, surrounded by an  anticoincidence arrangement. The effective area of the detector was 45 cm$^2$ with a FWHM of $1.4^{\circ}\times11^{\circ}$, X-rays in the energy range 17--115 keV were detected \cite{Ricker_et _al._1973}. Upper limit of the high energy  X-ray intensity for several sources.

\item \textbf{June 21, 1974} The high-energy x-ray telescope consisted of a 740 cm$^2$ phoswich type x-ray detector, sensitive in the region 20--150 keV with two slat collimators $6^{\circ}\times6^{\circ}$ and $3^{\circ}\times3^{\circ}$ field of view (FWHM) respectively. During this flight were observed the Coma and the Perseus clusters as well as the Crab Nebula \cite{Sheepmaker_et_al._1974}.

\item \textbf{August 13, 1974} X-ray measurement during the lunar occultation of the Crab. The detector had an effective area of 700 cm$^2$, a $6^{\circ}\times6^{\circ}$ FWHM and was sensitive in the energy range 20--150 keV \cite{Ricker_et_al._1975}. 

\item \textbf{June 1, 1975} from Palestine. The balloon took onboard phoswich detectors with a plastic anticoincidence shielding, the instrumentation was complited by two collimators with FWHM of $3^{\circ}\times3^{\circ}$ and $1.5^{\circ}\times6^{\circ}$. The observations were made in the energy band 20--150 keV. During the flight the source Cyg X-3 was observed and an energy-dependent light curve was advanced \cite{Ricker_et_al._1976a}. The transient X-ray source A0535+26 was also observed \cite{Ricker_et_al._1976b}.

\end{enumerate}

\newpage

\section{Appendix 3. List of the Balloon missions launched by world-wide institution.}

\begin{enumerate}
\item \textbf{April 13, 1965} Aire sur l'Adour (France). Bologna group GIFCO of the National research committee (CNR) and Bologna University. The experimental apparatus was a NaI(Tl) scintillator with an effective area of 5 cm$^2$, an effective FWHM of $\pm16^\circ$, and an energy interval that ranged between 20 and 200 keV \cite{Brini_et_al._1965}.

\item \textbf{June 18, 1965} Peterson and Jacobson (San Diego University) Crab nebula + Sco X-1. The detector consisted of a thin NaI scintillation counter surrounded by a 2 cm thick cylindrical CsI scintillation shield connected in electrical anticoincidence. The detector area was 9.2 cm$^2$. Observations were made in the energy range 10--200 keV \cite{Peterson_and_Jacobson_1966}.

\item \textbf{Marz  31, and April 2, 1966} Aire sur l'Adour (France). Bologna group GIFCO of the National research committee (CNR) and Bologna University.  Cyg X-2 and Ser X-1. NaI(Tl) scintillation counter, effective area 49 cm$^2$, energy range 20--180 keV. \cite{Brini_et_al._1967}

\item \textbf{April 5, 1966} from De Bilt, the Netherlands. The Dutch group led by Johan Bleeker made observation of the sources in the Cygnus region to determine their spectra \cite{Bleeker_et_al._1967}

\item \textbf{July-August 1966} Two balloon flights with identical X-ray detectors were carried out in the summer of 1966, one from De Bilt, the Netherlands (geomagnetic latitude 53$^\circ$ N), and the other from Taiyomura, Japan (geomagnetic latitude 25$^\circ$ N). The detector consists of a NaI(Tl) crystal, 12.5 mm thick and 5 mm in diameter, surrounded by an effective collimator-shield and a plastic scintillator guard counter. The rotating disk incorporated enables the separation of forward X-rays from the cosmic-ray-induced background. The results of the flights are in very good agreement with each other. In view of the rather large difference in geomagnetic latitude in these two flights, this agreement supports the celestial origin of the primary X rays observed. The energy spectrum between 20 and 180 keV can be expressed by a power law \cite{Bleeker_et_al._1968}
\item
\textbf{April 20, 1967}, flight No.54 Centre d'Etudes Nucléaires Saclay (France). The payload consisted of a gondola azimuthally stabilised with respect to the earth's magnetic field carrying a NaI (Tl) scintillation detector of effective area 95 m$^2$ and efficiency close to 1 between 15 and 150 keV. The photomultiplier pulses were transmitted to the ground after coding, as well as the data relative to the pressure and the attitude of the gondola. A collimator limited the field of view to 30$^\circ$ in the direction parallel to the North-South axis and 4$^\circ$ 30$^\prime$ in the perpendicular axis.
During the flight were observed the region of the sky which includes the 4 sources discovered in the Cygnus Constellation as X-rays emitters in the 1 to 10 keV range \cite{Rocchia_et_al:_1969}.


\item \textbf{Jun 4, 1967} from Palestine, Rice University Houston (Texas). The instrument flown, in this and in the following two experiments of the Rice University group, consisted of an NaI(Tl) scintillator of 10 cm diameter and 5 cm thickness. It was surrounded by another NaI(Tl) crystal 30 cm long and 25.4 cm in outer diameter that acted as a collimator. A thin (0.64 g/cm$^2$) plastic scintillator covered the 75 cm$^2$ aperture. The instrument was sensible in the energy range 35 to 560 keV  \cite{Haymes_et_al._1968a}, \cite{Haymes_and_Craddock_1966}. Gamma radiation from the Crab Nebula (the X-ray and gamma ray regions of the spectrum are characterized by the same index) was observed. Successive analysis of the data lead to the  "Detection of repetitive pulses of hard X-rays and gamma rays in data obtained from the Crab Nebula. The pulses are believed to originate from NP 0532" \cite{Fishman_et_al._1969}.

\item
\textbf{June 29, 1967}, flight No.84 from Aire sur l'Adoure by Centre d'Etudes Nucléaires Saclay (France). The goal of this launch was the observation of the source Tau X-1 in the region of the Crab Nebula. The payload was the same of the flight No. 54 of the previous flight on April the 20th. (s. above) \cite{Rocchia_et_al:_1969}

\item \textbf{August 10, 1967} from Palestine, Rice University Houston (Texas). Observation of the Virgo Region, probably candidate as X-ray source the elliptic galaxy M87 or the quasar  2C273. The payload consisted of the same detector as the previous mission on the 4th June 1967 (as reported above) \cite{Haymes_et_al._1968b}.

\item
\textbf{August 20, 1967}, flight No.92 Centre d'Etudes Nucléaires Saclay (France). Same payload as the flight No.54 of the 20 April 1967 (s. above). This flight was performed to confirm the existence of the source  Cyg X-3 \cite{Rocchia_et_al:_1969}.

\item \textbf{August 29, 1967} ****Gamma Ray Experiment?****** from Palestine, Rice University Houston (Texas). The payload consisted of the same detector as the previous mission on the 4th June and 29th August 1967 (as reported above). A study of the spectrum of the emission from the Cygnus region was performed \cite{Haymes_et_al._1968c}

\item \textbf{Februar 9, 1968} Adelaide University from Mildura (Australia). Spectral Properties of the X-Ray Objects GX 3+1, GX 354-5 and Sco XR-1 \cite{Buselli_et_al._1968}, \cite{McCracken_1969}. The observing platform consisted of two independent detectors:  a NaI(Tl) crystal 2 mm thick, shielded by a 1-inch-thick NaI(Tl) well surmounted by a CsI(Na) crystal of 8.9$^\circ$ FWHM and effective area of 54 cm$^2$; and a CsI scintillator crystal of 19 cm diameter and 3 mm thick with an exposure area of 220 cm$^2$, this was shielded by an outer passive lead shield and an inner active plastic scintillator, a copper collimator of 10$^{\circ}\times30^{\circ}$ FWHM was posed in front of the telescope \cite{Buselli_1968}, \cite{Davison_et_al._1971}. 

\item \textbf{April 28, 1968} Flight from Hyderabad performed by the TATA Institute of Mumbai (India). "The X-ray telescope consisted on a NaI(Tl) crystal of effective area 97.3 cm$^2$ and thickness 4 mm optically coupled to a 5 inches of diameter photomultiplier tube. The crystal was surrounded by both active and passive collimators. The passive collimator was composed of a cylindrical graded shield of lead, tin and copper and the active collimator was a cylindrical plastic scintillator. The field of view of the telescope was 18.6$^\circ$ at FWHM. The energy range extended from 17 to 124 keV."\cite{Agrawal_et_al._1971}. The study of the spectra of different X-ray sources was conducted \cite{Agrawal_et_al._1970a}.

\item \textbf{October 2, and 9, 1968} Aire sur l'Adour France. No X-rays Flux from some Quasars. In both flights the experimental apparatus consisted of two identical NaI(Tl) crystals surrounded by plastic scintillators in anticoincidence mode. The total effective area was 130 cm$^2$, collimators provided a triangular response of $\sim$14 FWHM. The measurements were done in energy range 20--200 keV  \cite{Brini_et_al._1970}.

\item \textbf{December 22, 1968}  from Hyderabat, TATA Institute Mumbai (India). The X-ray telescope was the same flown on the 28th of April of the same year. The important result of the mission was the discovery of the sudden change in the intensity of the X-ray emission from Sco X-1 \cite{Agrawal_et_al._1969}.

\item \textbf{April 16, 1969} TATA Institute Mumbai (India). With the same detector as in the precedent balloons flown by the Indian group, new studies on the spectrum of Sco X-1 were conducted \cite{Agrawal_et_al._1970b}.

\item \textbf{May 12, 1969} from Palestine Texas. Study of the emission of some X-ray sources \cite{Webber_and_Reinert_1970}, \cite{Reinert_et_al._1970}, \cite{NCAR_Annual_Report_1969}.

\item \textbf{August 25, 1969} Adelaide University from Mildura. By means of the data registered during this flight the Australian research group intended to confirm the existence of a new X-ray source hypothesized by Conner, Evans and Belian \cite{Conner_et_al._1969}. They confirmed the existence of a source, however it wasn't possible to determine the exact position and associate it to any constellation. The detector flown was the same used in the previous balloon flights of the Adelaide University  group \cite{Davison_et_al._1971}.
 
\item \textbf{January 28, 1970} TATA Institute from Hyderabad (India). Using the data registered during this and the precedent three mission of the group (28 April 1968, 22 December 1968, 16 April 1969) the spectrum of diffuse cosmic X-ray background was studied \cite{Manchanda_et_al._1972}. The detector on board of this fourth mission was the same as that on the previous flights (see above \cite{Agrawal_et_al._1971}).  
  
\item \textbf{September 22, 1972} Sankiro balloon centre. experiment conducted by Nagoya University group. The balloon carried onboard an instrument for the detection of X-ray in the range 20-120 keV that consisted of two NaI(Tl) crystal 3mm thick and with an effective area of 78 cm$^2$ each. Two different collimators were used one of 20$^{\circ}\times20^{\circ}$, and the other of 40$^{\circ}\times40^{\circ}$. The mission studied the x-ray background spectrum \cite{Makino_1975}.


\item \textbf{July 6, 1974} Gap (France). Study on Perseus hard X-ray emission by Brini and Frontera group in Italy. The measurement was taken by means of a NaI detector with an effective area of 300 cm$^2$ and a triangular response of 13$^\circ$ FWHM due to semi active collimators. The detector was sensitive in the range 20--210 keV \cite{Brini_et_al._1976}
\end{enumerate}

\section{\textit{Appendix 4. Ballons flown by AIT and MPI.}}

\begin{enumerate}
\item 
\textbf{August 28, 1973} from Palestine Texas, the  Crab-Nebula and the Crab-Pulsar has been observed at high energy ($>$30~keV)\cite{Arnold_et_al._1974}.

\item 
\textbf{February 20, 1975} from Palestine Texas a 4.8 h intensity variation of Cyg~X-3 \cite{Pietsch_et_al._1976} and the low state of the black-hole candidate Cyg~X-1 \cite{Steinle_et_al._1982} have been detected.

\item 
\textbf{May 3, 1976} from Palestine, Texas. The 1.24 s pulsations and the cyclotron line from Her X-1 were important results obtained with this experiment \cite{Truemper_et_al._1978}.

\item 
\textbf{September 3, 1977}, from Palestine. The  spectral feature in the  1.24 s pulsations was confirmed with large significance ($>$ 10 $\sigma$) by Her X-1 spectrum \cite{Voges_et_al._1982}.

\item 
\textbf{September 20, 1977} from Palestine. during this experiment X-ray binary containing a white dwarf, AM Hercules, has been observed  with a phoswich-type scintillator telescope of 766~cm$^2$ \cite{Staubert_et_al._1978}

\item 
\textbf{October 18, 1977} during the Palestina´s AIT/MPE  balloon observation the  Hard X-ray emission (20--90 keV) from Cyg~X-3 was measured with high statistical accuracy \cite{Reppin_et_al._1979}.

\item 
\textbf{November 22, and 24, 1978} Alice Spring, Australia. Observation of the X-ray binary  4U~1700-37/HD 153919 in the energy range 20--180~keV \cite{Pietsch_et_al._1980} and of the 283 s pulsation from 4U~0900-40 (Vela~X-1) at high X-ray energy ($\geq$ 18~keV) \cite{Staubert_et_al._1980}.

\item
\textbf{May 9, 1980} from Palestina the AIT and MPI group operated the first of a total of three successful balloon campaigns (other flight are listed below).  The balloons carried onboard the High Energy X-Ray Experiment (HEXE).
The scientific payload was divided into four detector bays, three of which contain the independent 800~cm$^2$ Phoswich scintillator detectors whilst the other bay holds the 114~cm$^2$ solid state detector assembly \cite{Proctor_et_al._1982}. The main targets of observations \cite{Staubert_and_Truemper_1982} were the neutron star Hercules~X-1, Cygnus~X-1, Sco~X-1 \cite{Jain_et_al._1984} the star HZ Herculis,  pulsating X-ray double stars, the Crab nebula and pulsar, and the extra-galactic object Cen~A. 

\item 
\textbf{September 28, 1981} Second flight of the HEXE experiment from Palestine. The scientific payload was the same as in the previous flight, namely a Phoswich scintillator detector and a 50 mm thick CsI(T1) crystal of 664 cm$^2$ effective area and a FWHM of $
\sim 3.5^\circ$. 

\item
\textbf{November 25, 1982} The third launch of the AIT-MPI group took place from Uberarba, Brazil. Unfortunately this mission failed due to an hole in the balloon and no scientific measurements have been possible.
\end{enumerate}
\newpage 

\section{Appendix 5. Transatlantic Balloons}

\begin{enumerate}

\item
\textbf{July 29, 1976} HXR--76 acronym of Hard X-Rays 1976 was the first transatlantic mission performed using a multi-instrumented gondola with several experiments  in the domain of Hard X-Rays, Gamma rays, Infrared Astronomy and biology aported by Italian Universities and Research Institutes exclusively\cite{stratocat_hrx76}developed by the Laboratory TESRE in Bologna, was composed by two identical and independent hard X-ray detectors  made up of a cylindrical NaI(Tl) crystal, optically coupled to three photomultipliers. Semiactive collimation, obtained by Pb shields surrounded by anticoincidence plastics scintillators gave a triangular field of view of 14 degreees FWHM. The effective area of the full system was 525~cm$^2$  while the  energy range was 20--300~keV. 
Another experiment, developed by the Laboratorio IAS, Frascati, comprised a large-area spectroscopic proportional chamber of 800 cm$^2$ filled with high-pressure xenon (at 4 atmospheres)  having energy range 25--190~keV. A square cells collimator limited the field of view to 14$^\circ$ FWHM \cite{Costa_et_al._1978}. 
Additional experiments were also provided by Istituto di Ricerca sulle Onde Elettromagnetiche (IROE) (Florence), IFCAI (Palermo) plus a few small piggy-back instruments. 
The balloon was launched  on July 29, 1976. In spite of the low spectral resolution and sensitivity a number of good scientific results have been obtained. Manly the first detection of X-ray emission above 80~keV from the Seyfert galaxy NGC 4151\cite{Ubertini_1981}.

\item 
\textbf{August 26, 1979} HXR--79 or Celimene II experiment was flown during august 1979. The scientific payload consisted of two \textit{}{second generation} high pressure Multiwire Spectroscopic Proportional Counters (MWSPC) with 900 cm$^2$ sensitive geometric area each, that observed in the hard X-Ray domain (16--180 keV). The field of view of the detectors was limited by means of two graded passive collimator having $7.5^\prime\times7.5^\prime$ and $5^\prime\times5^\prime$ FWHM aperture respectively. The detection of a 73 keV emission line in spectrum of the Crab Nebula was reported by \cite{Ubertini_et_al._1980,Polcaro_et_al._1980,Manchanda_et_al._1982}.
\item 
\textbf{August 10, 1980}
HXR--80M or Circe experiment, designed and build by IAS team,  was launched on August 10, 1980. Two large area MWPC units with  sensitive area of 2700 cm$^2$ each and spectral resolution of 13\% at 60~keV were integrated on a stabilized gondola, looking at the zenit. The detection of Hard X-rays emission (15--150 keV) from the region of 4U O515+38 has been reported by \cite{Ubertini_et_al._1982}.

\item 
\textbf{July 21, 1981}
HXR81--M or POKER experiment has been successfully flown during summer 1981.  The payload basically consists of four passively collimated MWPC’s similar to HXR80M experiment (geometric sensitive area of 2700~cm$^2$ each with an efficiency higher than 10\% in the operative range 15--200 KeV) with a square field of view of $5^\circ\times5^\circ$  FWHM. The overall spectral resolution was 13\% at 60 keV. The four MWPC’s were filled with a very pure gas mixture of Xenon-Argon-Isobuthane at high pressure (3 bars). 
In the survey 14 already identified  sources in the soft and/or  hard range, were detected with  higher than 4$\sigma$ statistical significance\cite{Ubertini_et_al._1983}. 





\end{enumerate}

\end{document}